\documentclass[a4paper,11pt]{article}

\usepackage[top=3.7cm,bottom=3.7cm,left=3cm,right=3cm]{geometry}
\usepackage[pdftex]{graphicx} 
\usepackage{amssymb,amsmath,amsthm,amsfonts}
\usepackage{xspace}
\usepackage{tabularx}
\usepackage{indentfirst}
\usepackage{subfigure}
\usepackage[small]{caption}
\usepackage{eucal}
\usepackage{eso-pic}
\usepackage{url}
\usepackage{booktabs}
\usepackage{afterpage}
\usepackage{parskip}
\usepackage{listings}
\usepackage{fancyhdr}
\usepackage{textcomp}
\usepackage{cite}
\usepackage{multirow}
\usepackage[utf8]{inputenc}   
\usepackage[english]{babel}   
\usepackage{setspace}
\usepackage{verbatim}
\usepackage{braket}
\usepackage[hidelinks]{hyperref}
\usepackage[normalem]{ulem}
\usepackage{cleveref}
\usepackage{tikz}
\usepackage{appendix}

\def\Mp9{M_{\mathrm{p},9}}

\begin{document}

\numberwithin{equation}{section}

\begin{flushright}
\end{flushright}
\pagestyle{empty}

\vspace{20mm}

 \begin{center}
{\LARGE \textbf{{$E_9$ symmetry in the heterotic string on $S^1$ \\ and the weak gravity conjecture}}}

\vspace{10mm} {\Large Veronica Collazuol, Mariana Gra\~na and Alvaro Herr\'aez}\\

\vspace{5mm}

Institut de Physique Th\'eorique, \\
Universit\'e Paris Saclay, CEA, CNRS, \\
Orme des Merisiers,  F-91191 Gif sur Yvette, France \\

\vspace{.5mm} {\small\upshape\ttfamily veronica.collazuol, mariana.grana, alvaro.herraezescudero @ ipht.fr} \\

\vspace{15mm}

\textbf{Abstract}
\end{center}

\vspace{3mm}

We show that compactifications of the heterotic string on a circle exhibit at the boundary of moduli space ($R\to 0$, or equivalently the decompactification limit $R \to \infty$) a tower of winding or momentum modes that enhance the $E_8 \times E_8$ or $SO(32)$ gauge symmetry to the affine algebras $(E_9 \oplus E_9)/\sim$ (the identification means that the two copies of $E_9$ share the same central extension) and $\hat{D}_{16}$, respectively. We also prove that these towers of modes satisfy the lattice weak gravity and repulsive force conjectures.

\newpage
\setcounter{page}{1}
	\pagestyle{plain}
	
\tableofcontents

\section{Introduction}
One of the most interesting results on the Swampland Program \cite{Vafa:2005ui} (see also \cite{Brennan:2017rbf,palti,vanBeest:2021lhn,Grana:2021zvf,Harlow:2022gzl} for some recent reviews), which aims to separate the low energy theories that can be consistently coupled to Quantum Gravity from those that cannot, is the presence of infinite towers of states that become light at the boundaries of moduli space. This feature of Quantum Gravity theories was originally suggested to be universal in \cite{Ooguri:2006in} with the formulation of the \emph{Swampland Distance Conjecture} (SDC). 

 Special points of moduli space of string theory compactifications, where symmetry enhancements can occur, are of particular interest in this context. 
Furthermore, an exhaustive scan of these points is also crucial for the question of string universality and whether the string lamppost can cover all viable theories of Quantum Gravity. This has recently been explored in the context of heterotic compactifications \cite{1,5,Font:2021uyw,Fraiman:2021soq}, F-theory constructions \cite{SZ,Hamada:2021bbz,Lee:2021qkx,Lee:2021usk} and also from general supergravity and swampland arguments \cite{Montero:2020icj,Hamada:2021bbz,Tarazi:2021duw, Long:2021jlv}.  
A subset of special points lie at infinite distances, where the tower of massless modes predicted by the SDC fits very naturally with the idea of a symmetry enhancement. 

A particularly simple, yet very interesting boundary of moduli space, is the $R\to 0$ (or its dual $R \to \infty$) limit of the compactification of the heterotic string on a circle. As in the bosonic string, one gets a tower of winding (momentum) modes that becomes massless in this limit, in the form proposed by the SDC.  The intriguing aspect in the heterotic string is that, as we will show, a subset of this tower enhances the $E_8 \times E_8 \times U(1)$ and $SO(32) \times U(1)$ gauge symmetries of the nine-dimensional theory (with vanishing Wilson lines) to the affine Lie algebras $(E_9 \oplus E_9)/\sim$ (where the identification $\sim$ means that the two copies of $E_9$ share the same central extension) and $\hat D_{16}$, respectively. For compactifications on $T^2$, this appearance of loop algebras in the dual F-theory on K3 was analysed first in \cite{DeWolfe:1998pr} and recently studied in great detail in \cite{Lee:2021qkx, Lee:2021usk}. Our results from the heterotic perspective match the expectations from these results obtained in the F-theory picture.

The appearance of infinite-dimensional symmetries in the circle compactification of the heterotic string can be anticipated from the Extended Dynkin diagram of the Narain lattice of states $\Gamma_{17,1}$ \cite{GoddardOlive1985}.  This 19-node Dynkin diagram contains, for the $E_8 \times E_8$ heterotic string, two copies of the affine algebra $E_9$, where the affine nodes are charged with respect to the circle, connected by a central node. By deleting nodes in this diagram one can obtain all the enhancement groups that arise in the circle compactifications of the heterotic string, as well as the point or subspace in moduli space where they arise \cite{1,5}. This diagram tells us that an enhancement to $(E_9 \oplus E_9)/\sim$ is possible with zero Wilson lines at the boundary of moduli space, either at $R \to 0$ if the affine nodes have winding charge, or in the decompactification limit $R \to \infty$ in the dual conventions, where these nodes have momentum charge. We  show that the OPE algebra of the states that become massless in this limit is indeed that of $(E_9 \oplus E_9)/\sim$ (or $\hat D_{16}$ if one starts with the $SO(32)$ heterotic string).  

It is interesting to see how the tower of states fits within the \emph{(Lattice) Weak Gravity Conjecture} (LWGC) \cite{wgc,lwgc,slwgc}, the closely related \emph{(Lattice) Repulsive Force Conjecture} (LRFC) \cite{Palti:2017elp,3}, and the SDC. The relation between the SDC and the WGC/RFC at the infinite distance points in moduli space at which a gauge couplings goes to zero has been studied in the literature \cite{Grimm:2018ohb,weaklimit, Gendler:2020dfp}, as it allows to connect the tower of light particles predicted by the former with the charged states becoming light as the gauge coupling decreases, predicted by the latter. In our case, the LWGC/LRFC require the existence of a superextremal (self-repulsive) state at every point in the (infinite) charge lattice of the dimensionally reduced theory, which includes the winding and KK gauge groups. By going to one of the aforementioned infinite distance limits, it can be seen that indeed some of the gauge couplings tend to zero, hence forcing the infinite number of states in the subset of the charge lattice corresponding to those gauge couplings to become lighter and lighter. We explicitly show that a subset of these states predicted by the LWGC/LRFC, namely the BPS vectors after compactification, are the ones that form the $(E_9 \oplus E_9)/\sim$ algebra. 

The paper is organised as follows. In Section \ref{sect1} we briefly present the heterotic string on $S^1$ to set our notation and introduce the Extended Dynkin Diagram (EDD) of the Narain lattice of the compactification, relating it to the study of symmetry enhancements at finite distance in moduli space. In Section \ref{sect2} we show how this tool can be applied also to the scan of the boundary of the heterotic moduli space for the $S^1$ compactification, and explicitly build the algebra of the vectors that are asymptotically massless. The detailed computations are presented in Appendix \ref{app:convention}.
After briefly reviewing the most relevant conjectures that concern our setup at the beginning of Section \ref{sect3}, we study their interplay with the algebra of vectors becoming massless in the infinite distance limits found in the previous section. The detailed analysis and the computations supporting our discussion are summarized in Appendix \ref{app:RFC}. 
Finally, we summarize and discuss our results in Section \ref{conclusions}.

\section{Compactifications of the heterotic string on $S^1$ and the Extended Dynkin Diagram}
\label{sect1}
In this section we recall the main features of compactifications of the heterotic string on $S^1$. We concentrate on the $E_8 \times E_8$ heterotic string, but all the results can be straightforwardly translated to the $SO(32)$ case. The circle is parameterised by the (dimensionfull) coordinate  $x^9 \equiv y \sim y + 2 \pi \sqrt{\alpha'}$, with a constant background metric  $g_{99}=\frac{R^2}{\alpha'}$ and Wilson line $A^{\hat{I}}, \, \hat{I}=1,...,16$ (with mass dimension +1). From now on, we will take $\alpha'=1$. 

The relevant part of the world-sheet action reads
\begin{equation}
\label{eqn:action}
    S = - \frac{1}{4\pi}  \int d^2 \sigma \Big[ \eta^{\alpha \beta} \Big( g_{MN} \partial_{\alpha}X^{M}\partial_{\beta}X^{N} + \frac{1}{2} \delta_{\hat{I}\hat{J}}  \partial_{\alpha}X^{\hat{I}} \partial_{\beta}X^{\hat{J}} \Big) +  \epsilon^{\alpha \beta} A^{\hat{I}}\partial_{\alpha}Y\partial_{\beta}X^{\hat{I}}) \Big] \, ,
\end{equation}
where the index $M=0,...,9$. The left and right (dimensionless) internal momenta along the compact circle direction are given by
\begin{equation}
\label{eqn:1}
    p_{L,R}=\sqrt{\frac{1}{2}}\Big(n \pm wR^2 -  A^{\hat{I}} \pi^{\hat{I}} - \frac{w  |A|^2}{2} \Big) \, ,
\end{equation}
and on the 16 dimensional ``heterotic torus" they take the form
\begin{equation}
\label{eqn:2}
    p^{\hat{I}}= \pi^{\hat{I}} + w A^{\hat{I}} \, .
\end{equation}
Here $n$, $w$ and $\pi^{\hat{I}}$ refer to the quantized charges, namely the momentum number, the winding number and a vector in the even, self-dual root lattice of $E_8 \times E_8$, which we will denote $\Gamma_8 \times \Gamma_8$, respectively. Moreover, we have $|A|^2=A^{\hat{I}} A^{\hat{I}}$. 

For any value of the moduli, the momentum vector $\boldsymbol{p}=(p_R, p_L, p^{\hat{I}}) \equiv (p_R,\boldsymbol{p}_L)$ spans an even and self dual Lorentzian lattice  with signature $(1, 17)$ endowed with the scalar product
\begin{equation}
    \boldsymbol{p} \cdot \boldsymbol{p} \equiv \boldsymbol{p}_L^2 - p_R^2 = 2wn + |\pi|^2 \in 2 \mathbb{Z}\, . 
\end{equation}
This is the so called Narain lattice $\Gamma_{(1,17)}$. It is unique up to $SO(1,17; \mathbb{R})$ transformations, so that the moduli space of the theory is
\begin{equation}
    \frac{O(1,17;\mathbb{R})}{O(17) \times O(1,17;\mathbb{Z})} \, ,    
\end{equation}
where $O(1,17; \mathbb{Z})$ is the T-duality group, described in detail for example in \cite{1}. 
The mass of the NS states is
\begin{equation}
\label{eqn:3}
    M^2= \boldsymbol{p}_L^2 + p_R^2 + 2 \Big( N + \bar{N} -\frac{3}{2} \Big),
\end{equation}
$N$ and $\bar{N}$ being respectively the left and right oscillator numbers, which should satisfy the Level Matching Condition
\begin{equation}
\label{eqn:4}
    \boldsymbol{p}_L^2 - p_R^2 + 2 \Big( N - \bar{N} -\frac{1}{2} \Big) =0 \, .
\end{equation}
In particular, the states that are massless everywhere in moduli space are the $N=1, \bar{N}=\frac{1}{2}, \boldsymbol{p}_L=0, p_R=0$ states, which split according to their nine-dimensional indices giving rise to the following spectrum.
    \begin{itemize}
        \item The gravitational sector:
   \begin{equation}
                 \alpha^{\mu}_{-1}\bar{\psi}^{\nu}_{-\frac{1}{2}}\ket{0}_{NS} \, \longrightarrow\,  g_{\mu \nu}, B_{\mu \nu}, \Phi \, .
    \end{equation}
        \item The vector bosons:
    \begin{equation}
    \label{eqn:12}
          \alpha^{\mu}_{-1} \bar{\psi}^{9}_{-\frac{1}{2}} \ket{0}_{NS}\, , \   \alpha^{9}_{-1} \bar{\psi}^{\mu}_{-\frac{1}{2}} \ket{0}_{NS}\, , \    \alpha^{\hat{I}}_{-1} \bar{\psi}^{\mu}_{-\frac{1}{2}} \ket{0}_{NS} \, \longrightarrow \, (g_{\mu 9}\mp B_{\mu 9}), \,  A_{\mu}{}^{\hat{I}} \, ,
        \end{equation}
        giving in general a $U(1)^{17}_L \times U(1)_R$ symmetry.
        \item The scalars:
        \begin{equation}
           \alpha_{-1}^{\hat{I}}\bar{\psi}_{-\frac{1}{2}}^{9}\ket{0}_{NS}\, , \   \alpha_{-1}^{9}\bar{\psi}_{-\frac{1}{2}}^{9}\ket{0}_{NS} \, \longrightarrow \,  A^{\hat{I}} , \, g_{99} \, .
        \end{equation}
    \end{itemize}
By looking at the additional massless vectors, if any, at a given point in moduli space one can find the gauge symmetry of the theory at that point. All the finite dimensional symmetry enhancements were found in \cite{1}. 
At any enhancement point in moduli space, the massless gauge vectors are characterized by $p_R=0$, $\bar{N}=\frac{1}{2}$ and by momenta $\boldsymbol{p}$ which are the roots of a group $G_r$ of rank $r \leq 17$, which is a product of ADE groups. The gauge group of the theory is then $G_r \times U(1)_L^{17-r} \times U(1)_R$.
For example, in the case of vanishing Wilson lines  all the states with $N=0, \bar{N}=\frac{1}{2}$ and $\pi^{\hat{I}}\in \Gamma_8 \times \Gamma_8, \, |\pi|^2=2$ are massless for any value of $R$. In particular, the massless vectors
\begin{equation} \label{E8vectors}
     \bar{\psi}_{-\frac{1}{2}}^{\mu} \ket{0, \pi_\alpha}_{NS}\, \, \longrightarrow \, A_{\mu}^{\alpha}
\end{equation}
with $\alpha=1,...,480$ the roots of $E_8 \times E_8$, together with the ones in \eqref{eqn:12} enhance the symmetry to $E_8 \times E_8 \times U(1)_L \times U(1)_R$.\footnote{This is the case for all $R \ne 1$. If $R=1$  the symmetry is enhanced to $E_8 \times E_8 \times SU(2)_L \times U(1)_R$.} 

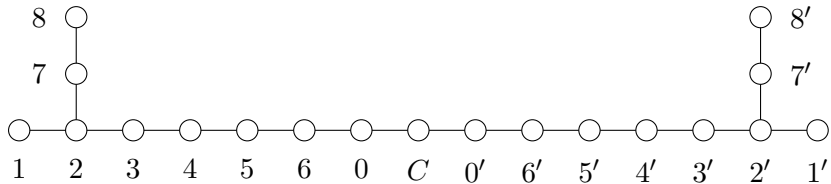
\begin{figure}[h!!]
 \centering
\begin{tikzpicture}

\draw(0,0)--(10.5,0);
\draw(0.75,0)--(0.75,1.5);
\draw(9.75,0)--(9.75,1.5);

\draw[fill=white](0,0) circle (0.14)node[below = 0.1 in]{$1$};
\draw[fill=white](0.75,0) circle (0.14)node[below = 0.1 in]{$2$};
\draw[fill=white](1.5,0) circle (0.14)node[below = 0.1 in]{$3$};
\draw[fill=white](2.25,0) circle (0.14)node[below = 0.1 in]{$4$};
\draw[fill=white](3,0) circle (0.14)node[below = 0.1 in]{$5$};
\draw[fill=white](3.75,0) circle (0.14)node[below = 0.1 in]{$6$};
\draw[fill=white](4.5,0) circle (0.14)node[below = 0.1 in]{$0$};
\draw[fill=white](5.25,0) circle
(0.14)node[below = 0.1 in]{$C$};
\draw[fill=white](6,0) circle (0.14)node[below = 0.1 in]{$0'$};
\draw[fill=white](6.75,0) circle (0.14)node[below = 0.1 in]{$6'$};
\draw[fill=white](7.5,0) circle (0.14)node[below = 0.1 in]{$5'$};
\draw[fill=white](8.25,0) circle (0.14)node[below = 0.1 in]{$4'$};
\draw[fill=white](9,0) circle (0.14)node[below = 0.1 in]{$3'$};
\draw[fill=white](9.75,0) circle (0.14)node[below = 0.1 in]{$2'$};
\draw[fill=white](10.5,0) circle (0.14)node[below = 0.1 in]{$1'$};

\draw[fill=white](0.75,0.75) circle (0.14)node[left = 0.1 in]{$7$};
\draw[fill=white](0.75,1.5) circle (0.14)node[left = 0.1 in]{$8$};

\draw[fill=white](9.75,0.75) circle (0.14)node[right = 0.1 in]{$7'$};
\draw[fill=white](9.75,1.5) circle (0.14)node[right = 0.1 in]{$8'$};

\end{tikzpicture}
      \caption{EDD of the $\Gamma_{(1,17)}$ lattice.}
       \label{fig:1}
\end{figure}

All the enhancement group algebras and the point in moduli space where they occur (up to T dualities) are encoded in the Extended Dynkin Diagram (EDD) of $\Gamma_{(1,17)}$ \cite{GoddardOlive1985} (as described for instance in \cite{2}), displayed in Figure \ref{fig:1}.

The labels to the nodes show how the $E_8 \times E_8$ lattice is embedded in $\Gamma_{(1,17)}$: the EDD is made of two extended $E_8$ Dynkin diagrams (adding the highest root 0 to the Dynkin diagram of each $E_8$) linked by a central node $C$. The primed indices are meant to distinguish the two copies of $E_8$. 
The nodes are described by a charge vector
\begin{equation}
\label{eqn:5}
    Z=(n,w,\pi^{i},\pi^{i'}) \in \Gamma_{(1,17)}  \, ,
\end{equation}
with $i=1,...,8$, as follows
\begin{equation}
\label{eqn:6}
    \begin{split}
    Z_i\, = & \, (0,0,\alpha_i,0_8)\, , \\
    Z_0\, =&\, (0,-1,\alpha_0,0_8)\, , \\
    Z_C\, =&\, (1,1,0_8,0_8)\, , \\
    Z_{0'}\, =&\, (0,-1,0_8,\alpha_{0'})\, , \\
    Z_{i'}\, =&\, (0,0,0_8,\alpha_{i'}) \, .
  \end{split}
\end{equation}
Here $\alpha_i$ are the simple roots of $E_8$ and $\alpha_0$ is the highest root, as explicitly shown in the Table \ref{tab:my_label}.
\begin{table}[h!!]
    \centering
    \begin{tabular}{|c|c|}
        \hline
         $ i $& $\alpha_i$  \\
         \hline
         1 & (1,-1,0,0,0,0,0,0) \\
         2 & (0,1,-1,0,0,0,0,0) \\
         3 & (0,0,1,-1,0,0,0,0) \\
         4 & (0,0,0,1,-1,0,0,0) \\
         5 & (0,0,0,0,1,-1,0,0) \\
         6 & (0,0,0,0,0,1,-1,0) \\
         7 & (-1,-1,0,0,0,0,0,0) \\
         8 & ($\frac{1}{2}, \frac{1}{2},\frac{1}{2}, \frac{1}{2},\frac{1}{2}, \frac{1}{2},\frac{1}{2}, \frac{1}{2}$) \\
         0 & (0,0,0,0,0,0,1,-1) \\
         \hline
        \end{tabular}
    \caption{Simple roots $(i=1,...,8)$ and highest root $(i=0)$ of $E_8$.}
    \label{tab:my_label}
\end{table}
\newline The primed quantities are taken to be equal to the unprimed ones, so they amount just to an identical copy of $E_8$, with the same convention.

With the following definition for the imaginary root
\begin{equation}
    \delta = (0, -1, 0_8, 0_8) \, , 
\end{equation}
the nodes 0 and 0' are embedded in $\Gamma_{(1,17)}$ respectively as
\begin{equation}
\label{eqn:share}
    (0, 0, \alpha_0, 0_8) + \delta  \, \text{ and } \,  (0, 0, 0_8, \alpha_{0'}) + \delta \, .
\end{equation}
This means that the extension of the two copies of the $E_8$ Dynkin diagram that build the EDD is performed with the same imaginary root.

We could have equivalently defined the charge vector as 
\begin{equation}
\label{eqn:7}
    Z=(w, n, \pi^i, \pi^{i'})
\end{equation}
instead of \eqref{eqn:5}, leaving \eqref{eqn:6} unchanged. In the first case, the nodes 0 and 0' have winding charge, in the second one momentum,  and the relation \eqref{eqn:share} still holds. The two cases are related by the T-duality transformation
\begin{equation}
\label{eqn:8}
    T_D= \begin{pmatrix}
        0 & 1 & 0 \\
        1 & 0 & 0 \\
        0 & 0 & \mathbb{I}
    \end{pmatrix} = \eta\, ,
\end{equation}
 where $\eta$ is the $O(1,17)$ invariant metric. This duality acts on $Z$ by exchanging momentum and winding, and on the background as
\begin{align}
\label{eqn:9}
    R'= & \frac{R}{R^2 + \frac{|A|^2}{2} }\, , \\ 
\label{eqn:10}
    A'^{\hat{I}}=& - \frac{A^{\hat{I}}}{R^2+\frac{|A|^2}{2}} \, ,
\end{align}
so that
\begin{equation}
    \mathcal{A}^{\hat I} \, \equiv \, \dfrac{A^{\hat I}}{R} \, =\,  - \dfrac{A'^{\hat I}}{R'}\,  \equiv \, - \mathcal{A}'^{\hat I} \, .
\end{equation}

All the possible enhancement groups, which are of the form $G_r \times U(1)^{17-r}$, are obtained by deleting $n=19-r$ nodes from the EDD in such a way to obtain the Dynkin diagram of the group $G_r$. The point or region in moduli space where this group arises is found by imposing that the remaining nodes satisfy the condition $\frac{p_R}{R}=0$. Each node gives a constraint on the moduli, that we collect in   
Table \ref{tab:fundamentaldomain} using conventions where the charge vector is given by \eqref{eqn:5}. As explained in \cite{2}, the co-dimension one plane in moduli space corresponding to each node is a fixed plane under a given T duality transformation.  Points of symmetry enhancement are therefore fixed points of T duality, and they arise at boundaries of a fundamental domain of moduli space. 
\begin{table}[h!!]
    \centering
    \begin{tabular}[h!]{|c|c|}
    \hline
         Node & Boundary of moduli space  \\
         \hline
        $ i$ & $\mathcal{A} \cdot (\alpha_i, 0_8) = 0$ \\
         0 & $ \mathcal{A} \cdot (\alpha_0, 0_8) = R \Big( 1 +\frac{|\mathcal{A}|^2}{2}\Big)$ \\
         C & $ R  \Big(\frac{1}{R^2} - \frac{|\mathcal{A}|^2}{2} \Big) = R $ \\
         0' & $\mathcal{A} \cdot (0_8, \alpha_{0'}) = R \Big( 1 +\frac{|\mathcal{A}|^2}{2}\Big)$ \\
         $i'$ & $\mathcal{A} \cdot (0_8, \alpha_{i'}) = 0 $  \\
         \hline
                 \end{tabular}
    \caption{Equations defining the co-dimension one boundary planes in moduli space corresponding to each node of the EDD, where the nodes 0 and 0', given in \eqref{eqn:6} have winding charge (i.e. $Z$ is that of Eq. \eqref{eqn:5}).}
    \label{tab:fundamentaldomain}
\end{table}
\newline Note that the assignment of the nodes with momentum and/or winding to a boundary in moduli space 
changes if we interchange momentum and winding. For the conventions given by \eqref{eqn:7}, the nodes given in \eqref{eqn:6} correspond instead to the boundaries given in Table \ref{tab:fundamentaldomain1}. The original assignment is more convenient for analysing the region of small $R$, while the latter is more suitable for large $R$. 
\begin{table}[h!]
    \centering
    \begin{tabular}[h!!]{|c|c|}
    \hline
         Node & Boundary of moduli space \\
         \hline
         $i$ & $\mathcal{A} \cdot (\alpha_i, 0_8) = 0$ \\
         0 & $ \mathcal{A} \cdot (\alpha_0, 0_8) = -\frac{1}{R}$ \\
         C & $R  \Big(\frac{1}{R^2} - \frac{|\mathcal{A}|^2}{2} \Big) = R $ \\
         0' & $\mathcal{A} \cdot (0_8, \alpha_{0'}) = -\frac{1}{R}$ \\
         $i'$ & $\mathcal{A} \cdot (0_8, \alpha_{i'}) = 0 $  \\
         \hline
                  \end{tabular}
    \caption{Equations defining the co-dimension one boundary planes in moduli space corresponding to each node of the EDD, where the nodes 0 and 0', given in \eqref{eqn:6} have momentum charge (i.e. $Z$ is that of Eq. \eqref{eqn:7}).}
    \label{tab:fundamentaldomain1}  
\end{table}

 The point or region in moduli space where enhancements of the form $G_r \times U(1)_L^{17-r} \times U(1)_R$ occur are found by imposing the $r$ equations given in Table \ref{tab:fundamentaldomain} for the remaining nodes in the EDD (or analogously one can use the conventions in Table \ref{tab:fundamentaldomain1}). If the node C is not deleted\footnote{It is easy to see that for finite ADE groups of maximal rank ($r=17$), node C cannot be deleted.}, these arise at a radius that is fixed in terms of the norm squared of the Wilson line, $R^2=1-\frac{|A|^2}{2}$.

\section{The $(E_9 \oplus E_9)/\sim$ enhancement}
\label{sect2}
Let us now consider what happens as we approach the limit $R\to \infty$. It can be seen that the following states become massless 
\begin{equation} \label{momtower}
    (0,n,\pi^{\alpha}), \quad n \in \mathbb{Z}, \, \pi^{\alpha} \in \Gamma_8 \times \Gamma_8 \, \text{or} \, \Gamma_{16}, \, |\pi|^2=2 \ .
\end{equation}
In the T dual case $R \to 0$, we get an equivalent tower of states becoming massless, labeled by the winding numbers
\begin{equation}
    (0, w, \pi^{\alpha}), \quad w \in \mathbb{Z}, \, \pi^{\alpha} \in \Gamma_8 \times \Gamma_8 \, \text{or} \, \Gamma_{16}, \, |\pi|^2=2 \, .
\end{equation}
Our goal now is to show that these states are actually the roots of an affine Lie algebra.

In the limit $R \to \infty$, and for any finite Wilson line $A$ (to be precise, this corresponds to the situation $\mathcal{A} \to 0$), the equations for all the nodes in Table \ref{tab:fundamentaldomain1}, apart from the one corresponding to node C, are satisfied. The corresponding algebra from the EDD is naively $E_9 \oplus E_9$. Taking into account the fact that the two copies of $E_9$ share the same imaginary root, we will denote the corresponding algebra as $E_9 \oplus E_9 / \sim$. As expected, there is also an enhancement to $E_9 \oplus E_9 / \sim$ at $R \to 0$ for zero Wilson line (namely $\mathcal{A} \to 0$) if we use the T-dual convention of Table \ref{tab:fundamentaldomain}, where the extended nodes 0 and 0' (c.f. eq. \eqref{eqn:6}) have momentum instead of winding charge. 

This pattern and the actual enhanced algebra can be understood by explicitly building the gauge vectors. For concreteness, we focus on the  $R \to \infty$ limit here, keeping in mind that the opposite limit is related to this one by T-duality. 
Moreover, all finite values for the Wilson lines $A^{\hat{I}}$ (including $A^{\hat{I}}=0$) are equivalent in the large $R$ limit since all of them correspond to the situation $\mathcal{A} \to 0$. For this reason, we approach the enhancement point along the path with vanishing Wilson line
\begin{equation}
\label{eqn:path}
    (A^{\hat{I}},R)=(0,R) \, ,
\end{equation}
but we remark that our results, namely the structure of the algebra, are the same even turning on a non-trivial (finite) $A^{\hat{I}}$, as discussed in detail in Appendix \ref{app:convention}. Note that along the path \eqref{eqn:path} the symmetry is $(E_8 \times E_8 \times U(1)_L) \times U(1)_R$ for all finite values of $R$. All these massless states are accompanied by a tower of momentum and winding states. We are interested in particular in the tower of momentum states of the $E_8 \times E_8$ vectors in \eqref{eqn:12} and \eqref{E8vectors}, given by 
        \begin{equation}
        \label{eqn:14}
         \alpha^{{\hat{I}}}_{-1} \bar{\psi}^{\mu}_{-\frac{1}{2}} \ket{0,n}_{NS}\, , \qquad    
        \bar{\psi}_{-\frac{1}{2}}^{\mu} \ket{0,n, \pi_\alpha}_{NS}\, ,
        \end{equation}
       where we recall that $\alpha=1,...,480$ and $\mu = 0, ..., 8$. 
These are associated to the following asymptotically conserved left currents 
\begin{equation}
\label{eqn:current}
    J^a_n(z) \equiv J^a(z) e^{i n y(z)} \, ,
\end{equation}
with $a=\{\hat{I}, \alpha\}$ and the finite algebra currents given by
\begin{equation}
\label{eqn:j}
    J^{\hat{I}}(z)= i \partial X^{\hat{I}}(z)\, , \qquad J^{\alpha}(z)= c_{\alpha} e^{i\pi^{\hat{I}}_{\alpha} X^{\hat{I}}(z)} \, .
\end{equation}
Here $c_{\alpha}$ are the cocycle factors, which give
\begin{equation}
    c_{\alpha} c_{\beta} \, = \,  \epsilon(\alpha, \beta) \, c_{\alpha + \beta}\, , \qquad \epsilon(\alpha,\beta) = \pm 1 
\end{equation}
when $\alpha+\beta$ is a root, and zero otherwise (and  $c_{0}\equiv 1$). 
The $J^a_n(z)$ have conformal dimension $h=1+\frac{n}{2R^2}$ which tends to 1 as $R \to \infty$, so indeed they correspond to conserved currents in the limit. By computing the OPEs of these currents we will prove that they  define an $(E_9 \oplus E_9)/\sim$ algebra. 

First, recall the OPEs of a set of currents  associated to an ADE algebra ${\cal G}$ at level 1, with conformal dimension $(h, \bar{h}) = (1,0)$, is given by
\begin{equation} \label{OPEgeneral}
    J^a(z) J^b(w) \sim \frac{1}{(z-w)^2} K^{ab} + i \frac{f^{ab}_{c}}{z-w} J^c(w) \ ,
\end{equation} 
where $K^{ab}$ is the Cartan-Killing metric and $f^{ab}_{c}$ are the structure constants. 
In the case of a finite algebra, eq. \eqref{OPEgeneral} is equivalent to the current algebra between the Laurent expansion coefficients defined by 
\begin{equation} \label{Jam}
    J^a_{k} = \oint \frac{dz}{2 \pi i} z^{k} J^{a}(z) \, , k \in \mathbb{Z} \, . \quad  
\end{equation}
This current algebra is thus given by
\begin{equation}
\label{eqn:curralg}
    [J^a_j, J^b_k] = j \delta^{ab} \delta_{j, -k} + i f^{ab}_{c} J^{c}_{j+k} \, ,
\end{equation}
and we can read the algebra ${\cal G}$, formed by the currents $J^{a}(z)$, from the algebra of the set $\{ J^{a}_0 \}$.

The structure constants of an $E_9$ algebra can be read off from the commutators 
\begin{equation}
\begin{split}    
\label{algE9}
    [H^I_n,H^J_m] &= \hat k\,  n \, \delta^{IJ} \delta_{n+m,0} \, , \\
    [H^I, E^{\alpha}_m] &= \alpha^I E^{\alpha}_{n+m} \, , \\
    [E_n^{\alpha}, E_m^{\beta}] &= 
    \begin{cases}
    N_{\alpha, \beta} E^{\alpha + \beta}_{n+m} \quad \quad \quad \quad \ \ \quad  \text{ for } \alpha+\beta \text{ root, } \\ 
    \alpha^I H^I_{n+m} + \hat{k}\,  n \, \delta_{n+m,0} \quad  \text{ for } \alpha=-\beta \, , \\
    0 \quad \quad \quad \quad \quad \quad \quad \quad \quad \ \ \, \text{ otherwise,}
    \end{cases} \\
   \end{split} 
\end{equation}
where $I=1,...8$, and $\hat k$ represents the central extension, which commutes with all the other generators. The $E_8$ roots $\alpha$ are normalized to $|\alpha|^2=2$, and $N_{\alpha, \beta}$ is a constant. 

Let us now turn to the calculation of the OPEs of the currents in \eqref{eqn:j} (see Appendix \ref{app:convention} for the detailed computation). For the towers of two Cartan currents we find 
\begin{equation}
\label{eqn:cc}
    J^{\hat{I}}_n(z) J^{\hat{J}}_m(w) \sim (z-w)^{\frac{nm}{2 R^2}} \Big( \delta^{\hat{I} \hat{J}} \frac{e^{i(n+m)y(w)}}{(z-w)^2} + i \delta^{\hat{I} \hat{J}} n \frac{:\partial y(w) e^{i (n+m)y(w)}: }{z-w} + \mathcal{O}(1) \Big) \, ,
\end{equation}
whereas the OPE of  the towers of two roots takes the form 
\begin{equation}
\label{eqn:rr}
    J^{\alpha}_n(z) J^{\beta}_m(w) \sim (z-w)^{\frac{nm}{2 R^2}} \cdot
    \begin{cases}
        \frac{\epsilon(\alpha, \beta) J^{\alpha+\beta}_{n+m}(w)}{z-w} + \mathcal{O}(1)\qquad \qquad \qquad \qquad \qquad \qquad\quad \ \alpha+\beta \,  {\rm root,} \\
        \frac{:e^{i (n+m) y(w)}:}{(z-w)^2} + \frac{\pi_{\alpha}^{\hat{I}} J^{\hat{I}}_{n+m}(w)+in:\partial y(w) e^{i (n+m)y(w)}:}{z-w} +\mathcal{O}(1) \, \, \, \, \alpha=-\beta\, , \\
        \mathcal{O}(1) \qquad \qquad \qquad \qquad \qquad \qquad \qquad\qquad \qquad \qquad {\rm \, otherwise.} 
    \end{cases}
\end{equation}
A Cartan and a root tower give
\begin{equation}
\label{eqn:cr}
    J^{\hat{I}}_n(z) J^{\alpha}_m(w) \sim (z-w)^{\frac{ nm}{2 R^2}} \Big( \frac{\pi^{\hat{I}}_{\alpha} J^{\alpha}_{n+m}(w)}{z-w} + \mathcal{O}(1) \Big) \, .
\end{equation}
Finally, the OPEs involving $\partial y$ are   
\begin{align} \label{JIdy}
    J^{\hat{I}}_n(z) i\partial y(w) \, \sim \, &   \frac{i}{2R^2} n \frac{:\partial X^{\hat{I}}(w) e^{i ny(w)}:}{z-w} + \mathcal{O}(1) \, ,\\
\label{Jalphady}
    J^{\alpha}_n(z) i\partial y(w) \, \sim \,  &   \frac{i}{2R^2} n \frac{:e^{i \pi_{\alpha}^{\hat{I}} X^{\hat{I}}(w)}e^{i n y(w)}:}{z-w} + \mathcal{O}(1) \, , \\
 \label{dydy}
    i\partial y(w) i\partial y(w) \, \sim \,  &  \frac{1}{2R^2} \frac{1}{(z-w)^2} + \mathcal{O}(1) \, .
\end{align}
The factors $(z-w)^{\frac{nm}{2 R^2}}$ in eqs. \eqref{eqn:cc}-\eqref{eqn:cr} tend to 1 in the limit $R \to \infty$, thus recovering the pole structure in \eqref{OPEgeneral}, which describes an algebra of (asymptotically, in this case) conserved currents. In this limit, all the OPEs involving the circle direction, namely \eqref{JIdy}-\eqref{dydy} vanish. 
Defining the generators as the zero modes of the asymptotically conserved currents ($a= \{ \hat I, \alpha \}$) 
\begin{equation}
\label{eqn:generators}
    (J^{a}_n)_0 \equiv \oint \frac{dz}{2 \pi i} J^{a}_n(z) \, ,
\end{equation}
and using the OPEs \eqref{eqn:cc}-\eqref{eqn:cr}, we obtain that they satisfy the following algebra (details of the computation can be found in Appendix \ref{app:convention})
\begin{equation}
\begin{split}
   & [(J^{\hat I}_n)_0, (J^{\hat J}_m)_0] = i n \delta^{\hat I \hat J} \delta_{n+m, 0} (\partial y)_0 \, ,\\
    &[(J^{\hat I}_n)_0, (J^{\alpha}_m)_0] = \pi_{\alpha}^{\hat I} (J^{\alpha}_{n+m})_0 \, ,\\
    &[(J^{\alpha}_n)_0, (J^{\beta}_m)_0] = \begin{cases}
          \epsilon(\alpha, \beta) (J^{\alpha + \beta}_{n+m})_0  \qquad \quad \qquad \quad \, \,    \,  \alpha+\beta \,  {\rm root}, \\
          \pi^{\hat I}_{\alpha} (J^{\hat I}_{n+m})_0 + i n \delta_{n+m, 0} (\partial y)_0 \quad \alpha=-\beta\, ,\\
         0 \qquad \qquad \qquad \qquad \qquad \qquad \, \,  \, {\rm otherwise,} 
    \end{cases}
    \end{split}
\end{equation}
where $(\partial y)_0$ corresponds to the zero mode of the Laurent expansion of $\partial y(z)$. 
Recall that all the commutators involving the zero mode of the circle direction vanish in the limit $R \to \infty$, as shown in \eqref{JIdy}-\eqref{dydy}. 

With a similar analysis, the non-vanishing components of the Cartan-Killing metric are found to be
\begin{equation}
   K(J^{\hat I}_n(z),J^{\hat J}_m(w))= \delta_{n+m,0} \delta^{\hat I \hat J} \, , \quad K(J^{\alpha}_n(z),J^{\beta}_{m}(w)) = \delta_{m+n,0} \delta^{\alpha+\beta,0} \, .
\end{equation}

These are precisely the commutators and the Cartan-Killing metric for two copies of the algebra $E_9$ with central extension given by 
\begin{equation}
    \hat k = (\partial y)_0 \ .
\end{equation}
Since both $E_9$ share the central extension, the total algebra is $(E_9\oplus E_9)/\sim$. The fact that the two $E_9$ share the central extension is consistent with the fact that the maximum rank of the left-moving algebra is 17. 
Notice that since we did not need to specify the roots of the algebra, the same applies to the $SO(32)$ case, which at the same boundary of moduli space gets enhanced to $\hat{D}_{16}$ (whose rank is 17 as well), in agreement with what one gets from the EDD with $SO(32)$ embedding (see \cite{5} for details).

In fact, it is known \cite{BPS} that the $E_9 \oplus E_9/\sim$ (or $\hat{D}_{16}$) states, together with the other BPS states which are Lorentz scalars in the left-moving sector and in the right-movers ground state, give rise a BPS algebra of the generalized Kac-Moody type even at finite distances in moduli space, where generically all of them are massive. The $(E_9 \oplus E_9)/\sim $ (or $\hat{D}_{16}$) algebra is then realized as a subalgebra at the boundary of the moduli space by the BPS vectors that become massless.

Note that $E_{10}$ (or $E_{11}$) cannot be realized in this setting, as this should appear at infinite distance in moduli space where clearly the equation for the node C can never be satisfied. Another interesting point to make is that even if the EDD suggests that it is possible to find groups of maximal enhancement in the left sector with only one copy of $E_9$, these cannot be realised in heterotic compactifications on the circle.  This can be seen either from Table 3, which says that for $R \to \infty$ the equations for both $E_9$ are satisfied at the same time, as well as from the explicit construction of the algebra, where the central extension extends both $E_8$ factors at the same time.

\section{Swampland conjectures approaching the $(E_9 \oplus E_9)/\sim$ infinite distance limit}
\label{sect3}
Let us now try to look at this symmetry enhancement from the point of view of the Swampland program \cite{Vafa:2005ui} (see also \cite{Brennan:2017rbf,palti,vanBeest:2021lhn,Grana:2021zvf,Harlow:2022gzl} for some recent reviews). As a full stringy construction, it is expected that our setup fulfills all the Swampland Conjectures, so it is interesting to see how the relevant ones are realized in this particular example. Moreover, it is particularly illuminating to see how constraining these conjectures are, in the sense of understanding what aspects about the enhancement could have been anticipated had we not known the full string construction, but some 9d EFT data instead.

On the one hand, it seems very natural to try to connect our setup to the 
\newtheorem*{SDC}{Swampland Distance Conjecture (SDC) \cite{Ooguri:2006in}}
\begin{SDC}
In a gravitational theory with a moduli space (with metric given by the kinetic terms of the scalars), starting at a point $P$ in such moduli space and moving towards a point $Q$ an infinite geodesic distance away, an infinite tower of states  becomes exponentially light (in Planck units) as
\begin{equation}
\dfrac{M_{\mathrm{tower}}(Q)}{M_{\mathrm{p}} (Q)} \sim \dfrac{M_{\mathrm{tower}}(P)}{M_{\mathrm{p}} (P)} \, e^{-\alpha d(P, Q)} \, ,
\end{equation}
where $M_{\mathrm{p}}(Q)$ is the EFT Planck Mass at the point $Q$ in moduli space, $d(P, Q)$ is the geodesic distance between the points $P$ and $Q$, and $\alpha$ is some positive, order-one number.
\end{SDC}
Since the points we have studied (i.e. $R\to 0, \ \infty$) are both infinite distance points, it is expected to find the corresponding towers when approaching them. In this case there is nothing unexpected, as these are just the winding and KK-states, respectively. It is  interesting though to analyze how these towers fit in with the tower versions of the Weak Gravity Conjecture (WGC)\cite{wgc} and the (related) Repulsive Force Conjecture (RFC)\cite{Palti:2017elp, 3}, given the fact that they are also charged under the other gauge groups that are present in the 9d theory. The relevant versions of such conjectures that we are interested in are the following\footnote{The currently accepted versions of these conjectures are the sub-Lattice WGC \cite{slwgc} and sub-Lattice RFC \cite{3}, since there are some examples in which they are not fulfilled for the full charge lattice but only a sub-lattice. However, since in our case they are fulfilled for the full charge lattice we refer to the lattice versions here.}
\newtheorem*{LWGC}{Lattice Weak Gravity Conjecture (LWGC) \cite{lwgc,slwgc}}
\begin{LWGC}
In the presence of multiple $U(1)$ gauge groups, for every point $\Vec{q}$ in the charge lattice $\Gamma$ of a theory which can be consistently coupled to quantum gravity there must exist a superextremal particle. 
\end{LWGC}
and
\newtheorem*{LRFC}{Lattice Repulsive Force Conjecture (LRFC) \cite{3}}
\begin{LRFC}
In order for a multiple-$U(1)$ gauge theory with charge lattice $\Gamma$ to be coupled consistently to quantum gravity, in any site $\Vec{q}\in \Gamma$ there exists a self-repulsive particle.  
\end{LRFC}

In the remaining of this section, we focus on the states charged under the $E_8\times E_8$ gauge bosons (more precisely under their Cartan generators), as well as those with winding and momentum charge. In the corresponding infinite distance limit, some subset of these gauge couplings go to zero and we show how the states predicted by the LWGC and the LRFC become massless, as required by the superextremality or self-repulsiveness conditions in the vanishing gauge coupling limit. These states in fact correspond to the $(E_9 \oplus E_9)/\sim$ vectors, therefore connecting the conjectures to the presence of the affine enhancement.
 
\subsection{The WGC, the RFC and the presence of $(E_9 \oplus E_9)/\sim$}

The extremality bound for black holes charged under the heterotic group in toroidal compactifications was computed in \cite{sen} and takes the form
\begin{equation}
\label{eq:extremalitybound}
    \frac{\alpha'}{4} M^2 \geq \frac{1}{2} \max(\boldsymbol{p}_L^2,p_R^2)\, ,
\end{equation}
so that states fulfilling the bound are subextremal. This is the relevant bound to study the superextremal particles predicted by the WGC.

On the other hand, we are going to compute the  long range interaction between probe particles charged under these gauge groups in the Einstein frame, which are the relevant quantities for the RFC. We use the compactified heterotic low energy action as in \cite{homh}, with the following metric ansatz 
\begin{equation}
\label{eqn:141}
    ds^2 = e^{\frac{4}{7} \tilde{\Phi}} e^{-\frac{2}{7}\tilde{\sigma}}g^{E}_{\mu \nu} dx^{\mu} dx^{\nu} + e^{2\tilde \sigma} (dy + Z_{\mu} dx^{\mu})^2, \, \, \, y \sim y + 2 \pi R \, ,
\end{equation}
where $\tilde{\Phi}$ and $\tilde{\sigma}$ denote the dynamical parts of the dilaton field, $\Phi=\Phi_0 + \tilde{\Phi}$, and the (dimensionless) radion, $\sigma=\sigma_0 + \tilde{\sigma}$, where we have also explicitly included a background piece for the scalars. We recall that the physical radius of the circle in the string frame and in string units is $R=e^{\sigma_0}\sqrt{\alpha'}$, and it is included in the definition of the coordinate $y$ instead of in the metric for convenience. The tensor $g^{E}_{\mu \nu}$ gives the 9 dimensional metric in the Einstein frame, and $Z_\mu$ represents the (dimensionless) graviphoton.

To compute the long range force we need the action up to quadratic order in the dynamical fields and including the minimal coupling to the physical states in the string spectrum, which act as semi-classical sources. We leave the details of the computation for Appendix \ref{app:RFC}, but outline the main points here. Using the metric ansatz in \eqref{eqn:141}, one needs to perform the following field redefinition to work with conventionally  normalized fields without mixed kinetic terms (as explained in detail around eq. \eqref{eqn:radionred}) 
\begin{equation}
    \lambda = \frac{4}{\sqrt{7}}\Big(\frac{\Phi}{4} - \sigma \Big) \, .
\end{equation}
Using this, the expression for the action up to quadratic order and including the minimally coupled sources that describe the heterotic states reads \begin{equation}
\label{eqn:15}
\begin{split}
    S \, =\, &\frac{\Mp9^7}{2} \int d^{9}x \sqrt{-g} \Big( R- \frac{1}{2} \partial_{\mu} \lambda \partial^{\mu} \lambda - \frac{1}{2} \partial_{\mu} \Phi \partial^{\mu} \Phi  -\frac{\alpha'}{2}  \partial_{\mu} A_{9}^{\hat{I}} \partial^{\mu} A^{\hat{I}}_9 - \frac{R^2}{4} \tilde{Z}_{\mu \nu} \tilde{Z}^{\mu \nu} -       \\
     & -  \frac{\alpha'}{4} \tilde{F}_{\mu \nu}^{\hat{I}} \tilde{F}^{\mu \nu \hat{I}}  -\frac{\alpha'^2}{4R^2} \tilde{W}_{\mu \nu} \tilde{W}^{\mu \nu} \Big) - \int M(\Phi, \lambda, A_9^I) \, ds - w \int \tilde{W} - \pi^{\hat{I}} \int A^{\hat{I}} - n \int \tilde{Z} \, ,
\end{split}
\end{equation}
where $ds$ is the line element along the particle world-line.
The 9 dimensional Planck mass has the following expression
\begin{equation}
\label{eqn:25}
    \frac{\Mp9^7}{2} = \frac{e^{-2\Phi_0}\, R}{(2\pi)^6 \alpha'^{4}}  \, , 
\end{equation}
whereas $\tilde{Z}_{\mu}$ is the dimensionfull graviphoton (see eq. \eqref{eqn:26}), and $A^{\hat{I}}_{\mu}$ and $\tilde{W}_{\mu} \equiv \boldsymbol{B}_{\mu 9}+ \frac{\alpha'}{2} A_{9}^{\hat{I}} A_{\mu \hat{I}}$ are the Cartan gauge fields, which are the only ones we are interested in to compute the long range force. All these gauge fields are normalized to have mass dimension +1 and in such a way that the charged states have integer charges, and $\tilde{Z}_{\mu \nu}, \, \tilde{W}_{\mu \nu} \text{ and } \tilde{F}_{\mu \nu}^{\hat{I}}$ are their respective field strengths. 

The repulsive force condition (for vanishing Wilson lines) is 
\begin{equation}
\begin{split}
    M^2 \leq  \Mp9^2 (32\pi^6)^{\frac{2}{7}}  \max \Big\{ & \left( 2|\pi|^2 e^{\frac{1}{2}\Phi_0}e^{\frac{\sqrt{7}}{14}\lambda_0} + n^2 e^{\frac{4 \sqrt{7}}{7}\lambda_0}+ w^2 e^{\Phi_0} e^{-\frac{3\sqrt{7}}{7} \sigma_0} + 2n w e^{\frac{1}{2}\Phi_0}e^{\frac{\sqrt{7}}{14}\lambda_0} \right) \, , \\
   & \left( n^2 e^{\frac{4 \sqrt{7}}{7}\lambda_0}+ w^2 e^{\Phi_0} e^{-\frac{3\sqrt{7}}{7} \sigma_0} - 2n w e^{\frac{1}{2}\Phi_0}e^{\frac{\sqrt{7}}{14}\lambda_0}\right) \Big\} \,.
     \end{split}
\end{equation}
Let us remark that this expression is equivalent to the extremality bound given in eq. \eqref{eq:extremalitybound}, so that all  superextremal states are self-repulsive. This means that the WGC and the RFC are equivalent in this case (as expected, see \cite{3}), but one must keep in mind that the long range force computation only makes sense in the perturbative regime.
The expressions of the three relevant gauge couplings as a function of the moduli and in Planck units are
\begin{align}
    \frac{1}{g_Z^2} \, =\,  & \frac{\Mp9^{5}}{2(32\pi^6)^{\frac{2}{7}}}  e^{-\frac{4\sqrt{7}}{7}\lambda_0}  \, , \\
    \frac{1}{g_W^2}\, =\, & \frac{\Mp9^{5}}{2(32\pi^6)^{\frac{2}{7}}} e^{\frac{3 \sqrt{7}}{7}\lambda_0}  e^{-\Phi_0} \, , \\
    \frac{1}{g_A^2}\, =\, & \frac{\Mp9^{5}}{2(32\pi^6)^{\frac{2}{7}}} e^{-\frac{\sqrt{7}}{14}\lambda_0}e^{-\frac{1}{2}\Phi_0}  \, .
\end{align}
Thus, in the infinite distance decompactification limit $e^{\lambda_0} \to  0$ ($\frac{R}{\sqrt{\alpha'}} \to \infty $)  we obtain in Planck units 
\begin{equation}
 e^{\lambda_0} \to 0 \, : \quad    g_Z\sim e^{\frac{2 \sqrt{7}}{7}\lambda_0} \to 0, \quad g_W\sim e^{\frac{-3 \sqrt{7}}{14}\lambda_0} \to \infty, \quad g_A \sim e^{\frac{\sqrt{7}}{28}\lambda_0} \to 0 \, .
\end{equation}
Since we are particularly interested in the states charged under the graviphoton and the heterotic gauge group (as these are the ones that form the $E_9 \oplus E_9/ \sim$ algebra), it is clear that the large radion limit is the right one, as the relevant gauge couplings are small and the perturbative calculation under control. Let us remark also that even though  the gauge theory associated to $W_{\mu}$ is non-perturbative in this corner of moduli space, we do not consider states charged under it here. We emphasize that the correspondence between the WGC and the RFC can be made precise in this limit, so that we can then interpret the presence of the $(E_9 \oplus E_9)/\sim$ gauge fields in the light of these conjectures. Restricting to the aforementioned charged states, we obtain the following superextremality/self-repulsiveness condition 
\begin{equation}
\label{eq:WGC/RFC bound}
    M \leq \Mp9^{\frac{7}{2}}  \sqrt{|\pi|^2 g_A^2 +\frac{n^2g_Z^2}{2}}\, .
\end{equation}
For finite gauge couplings, this bound allows for massive states. Moreover, since the states that become the $E_9 \oplus E_9/ \sim$ vectors in the infinite distance limit are BPS states, they saturate the inequality. In the decompactification limit, the fact that the two relevant gauge couplings tend to zero means that the LWGC/LRFC require an infinite tower of states (one for each point in the infinite charge lattice) to become massless. In particular, note that both the KK replicas of the Cartans and the roots of the $E_8 \times E_8$ heterotic group studied in the previous section, which have arbitrary $n$ and $|\pi|^2=0,2$  are forced to become massless. This means that the states of the $E_9 \oplus E_9/ \sim$ BPS algebra that we found in the limit can be understood to become massless as a consequence of the LWGC/LRFC. Note that in general there are many more (generically non-BPS) states that become massless in the limit according to \eqref{eq:WGC/RFC bound}, so that only a subset of all these asymptotically massless states, namely the BPS ones, form the $E_9 \oplus E_9/ \sim$ algebra. Moreover, following the reasoning in \cite{Grimm:2018ohb,Gendler:2020dfp}, we can argue for the appearance of this tower also as a way to prevent the restoration of a global symmetry, which is known to be forbidden in quantum gravity (see e.g. \cite{palti} and references therein). Since in the limit of vanishing coupling a gauge symmetry behaves as a global one, there must be a way to prevent this from happening at the infinite distance point, and this is indeed the case due to the presence of the infinite lattice of states becoming light. Note that this is also in complete agreement with the magnetic WGC if we identify the cutoff as the scale of the tower of states becoming light, as it would predict a cutoff (in Planck units) $\Lambda_{\mathrm{cutoff}}\lesssim g  \rightarrow 0$. Let us finally mention that the limit of vanishing radius would give analogous results upon application of T-duality, which is not directly manifest in the computations above as we are working in the Einstein frame instead of the string frame.

\section{Conclusions}
\label{conclusions}
In this work we studied the infinite distance points of the heterotic string compactified on $S^1$ that correspond to the decompactification limit and its T-dual vanishing size limit. We showed that as one approaches these limits, an affine algebra appears. In particular, we find that starting from the $E_8 \times E_8$  heterotic theory, the corresponding affine algebra is $(E_9 \oplus E_9)/\sim $ (the identification means that the two copies of $E_9$ share the same central extension), whereas starting from the $SO(32)$ theory, one gets $\hat{D}_{16}$. The central extension that gives rise to the affinization of these algebras is the circle direction. In fact, these heterotic BPS states which become arbitrarily light as we approach the infinite distance points are known to be part of an even bigger BPS algebra realized at finite distances, where BPS states are generically massive \cite{BPS}.

We motivated the appearance of these affine algebras by means of the Extended Dynkin Diagram of the Narain Lattice $\Gamma_{(1,17)}$, and also found them very explicitly by computing the relevant OPEs and taking the infinite distance limit (e.g. $R\to \infty$). To be precise, by computing the OPEs of the KK modes of the Cartan and root sectors of the $E_8\times E_8$ heterotic states, we were able to isolate the simple poles in the limit and read off the affine algebra, explicitly identifying its central extension. This explicit and clean statement is one of the main results of this work. Moreover, we find that no other affine algebras are obtained, that is we only find the affine versions of the algebras of $E_8 \times E_8$ and $SO(32)$. 

Infinite gauge groups  were also found in their dual F-theory reincarnation, at the boundaries of ``open string moduli space" \cite{DeWolfe:1998pr,Lee:2021qkx,Lee:2021usk}. In compactifications of F-theory on K3 one finds other affine exceptional algebras, and thus we expect to find them in toroidal compactifications of the heterotic string. We leave this for future work. We note, however, that our result that only the $(E_9 \oplus E_9)/\sim$ and $\hat{D}_{16}$ are found in the decompactification limit of the 9 dimensional theory is in agreement with the claim  that only the affine versions of the gauge groups realized in the decompactified theory (the 10 dimensional theory in our case) should be obtained in the decompactification limit of the lower dimensional one \cite{Lee:2021usk}.

We have also analyzed these results in the light of some relevant Swampland Conjectures, namely the Weak Gravity Conjecture, the closely related Repulsive Force Conjecture, and the Distance Conjecture. We have explicitly performed the field theory computation of the force between two probe heterotic states in the 9 dimensional theory, and showed that it matches the corresponding extremality bound \cite{sen} (as long as the perturbative computation can be trusted). The Lattice Weak Gravity/Repulsive Force Conjecture then requires one particle in each site of the charge lattice to be superextremal/self-repulsive, and in particular the BPS states saturate this bound. We showed explicitly that when the relevant infinite distance point is approached, a subset of the gauge couplings becomes vanishing (this has been conjectured to be a general property of all infinite distance limits in QG \cite{Gendler:2020dfp}), forcing the relevant states on the lattice to become massless. In the decompactification limit, these vanishing gauge copulings are associated to the heterotic gauge bosons (or more precisely, their Cartan subsector) and the graviphoton, and a subset of the states charged under them (i.e. the BPS ones) are precisely the ones that form the $(E_9 \oplus E_9)/\sim$ or $\hat{D}_{16}$ algebras. It would be interesting to study this interplay between the Lattice Weak Gravity/Repulsive Force Conjecture and the appearance of affine algebras, and its application to infinite distance points in more generality.

 \vspace{5mm}
{\bf \large Acknowledgments}

\noindent We would like to thank Martin Cederwall, Anamaria Font, Bernardo Fraiman, Ruben Minasian, Hector Parra De Freitas, Daniel Waldram and Timo Weigand for discussions, and particularly Carmen Nu\~nez and Peng Cheng for technical help and challenging discussions. This work was supported by the ERC Consolidator Grant 772408-Stringlandscape. 

\newpage
\appendix
\section{Details on the worldsheet realization of the affine algebras}
\label{app:convention}
In the following we are going to derive the $E_9 \oplus E_9 / \sim$ algebra from the CFT point of view in the general case of generic finite constant Wilson line $A$. 
With the normalization in \eqref{eqn:action}, the non trivial OPEs between the worldsheet scalars in the heterotic theory are
\begin{align}
\label{eqn:XX}
    X^{\mu}(z,\bar{z}) X^{\nu}(w,\bar{w}) \sim& - \frac{1}{2} \eta^{\mu \nu} \log|z-w|^2 \, , \\
\label{eqn:yyope}
    Y(z,\bar{z}) Y(w,\bar{w}) \sim &-  \frac{1}{2R^2} \log|z-w|^2 \, ,  \\
\label{eqn:oldII}
    X^{\hat{I}}(z) X^{\hat{J}}(w) \sim & - \delta^{\hat{I} \hat{J}} \log(z-w) \, ,
\end{align}
where $Y(z, \bar{z}) = y(z) + \bar{y}(\bar{z})\, $. The vertex operators in the 0 picture associated to each $E_8 \times E_8 \times U(1)$ (or $SO(32) \times U(1)$) massless vector in the string spectrum are
\begin{align}
\alpha^{{\hat{I}}}_{-1} \bar{\psi}^{\mu}_{-\frac{1}{2}} \ket{0,n}_{NS} \rightarrow  & \,  J^{\hat{I}}(z) \Big( i \sqrt{2} \bar{\partial} X^{\mu}(\bar{z}) + \frac{1}{\sqrt{2}} \kappa \cdot \bar{\psi}(\bar{z}) \bar{\psi}^{\mu}(\bar{z})\Big)e^{ikX(z, \bar{z})} e^{i n Y(z, \bar{z})} \, , \\
      \bar{\psi}_{-\frac{1}{2}}^{\mu} \ket{0,n, \pi_\alpha}_{NS} \rightarrow & \,  J^{\alpha}(z)\Big( i \sqrt{2} \bar{\partial} X^{\mu}(\bar{z}) + \frac{1}{\sqrt{2}} \kappa \cdot \bar{\psi}(\bar{z}) \bar{\psi}^{\mu}(\bar{z}) \Big)e^{ikX(z, \bar{z})}e^{i (n - \pi_{\alpha}^{\hat I} A^{\hat I}) Y(z, \bar{z})} \,  ,\\
    \alpha^{9}_{-1} \bar{\psi}^{\mu}_{-\frac{1}{2}} \ket{0}_{NS} \rightarrow  \, & J^9(z) \Big( i \sqrt{2} \bar{\partial} X^{\mu}(\bar{z}) + \frac{1}{\sqrt{2}} \kappa \cdot \bar{\psi}(\bar{z}) \bar{\psi}^{\mu}(\bar{z}) \Big)e^{ikX(z, \bar{z})}e^{i n Y(z, \bar{z})} \, .
\end{align}
They depend on the right momentum $\kappa=(k_{\mu}, p_R)$, on the $E_8 \times E_8$ (or $SO(32)$) currents, $J^{\hat I}$ and $J^{\alpha}$ as in \eqref{eqn:j} and on the $U(1)$ current 
\begin{equation}
    J^9(z) = i \sqrt{2} \partial y(z) \, .
\end{equation}
In the limit $R \to \infty$, in order to read the structure of the algebra in the presence of a non vanishing Wilson line we can generalize the definitions \eqref{eqn:current} of $\{ J^{a}_n \}$ ($a = \hat I, \alpha$) to 
\begin{equation}
\label{eqn:newcurrent}
\begin{split}
    J^{\hat I}_n(z)  &\equiv \Big(  J^{\hat I}(z) - \frac{A^{\hat I}}{\sqrt{2}} J^9(z) \Big) e^{i n y(z)} =  i (\partial X^{\hat I}(z) - A^{\hat I} \partial y(z)) e^{i n y(z)} \, , \\
     J^{\alpha}_{n}(z) & \equiv J^{\alpha}(z) e^{i(n - \pi_{\alpha}^{\hat I} A^{\hat I}) y(z)} = c_{\alpha} e^{i \pi^{\hat I}_{\alpha} (X^{\hat I}(z) - A^{\hat I} y(z)) }  e^{iny(z)} \, ,
\end{split}
\end{equation}
which are still both associated to massless states. They can still be interpreted as asymptotically conserved currents, as their conformal dimension ($h_{\hat I} = 1 + \frac{n}{2R^2}$ and $h_{\alpha}= 1 + \frac{(n-\pi_{\alpha}^{\hat I}A^{\hat I})}{2R^2}$ respectively) tends to 1 as $R \to \infty$. 
As one can see from \eqref{eqn:newcurrent}, they can be rewritten using the field redefinition 
\begin{equation}
    X'^{\hat I}(z) = X^{\hat I}(z) - A^{\hat I} y(z) \,
\end{equation}
(which makes it manifest that the effect of the Wilson line is to mix the $E_8 \times E_8$ Cartan states with the left KK one) yielding
\begin{equation}
    J^{\hat I}_n(z) = i \partial X'^{\hat I}(z) e^{iny(z)} \, , \qquad  J^{\alpha}_n(z) = c_{\alpha} e^{i \pi_{\alpha}^{\hat I} X'^{\hat I}} e^{iny(z)} \, ,
\end{equation}
which are in the same form as the ones in the zero Wilson line case. In terms of the redefined heterotic coordinate, the non trivial OPEs among the worldsheet fields are
\begin{align}
    X'^{\hat I}(z) X'^{\hat J}(w) \sim & X^{\hat I}(z) X^{\hat J}(w) + A^{\hat I} A^{\hat J} y(z) y(w) \sim - \Big( \delta^{\hat I \hat J} + \frac{A^{\hat I} A^{\hat J}}{2R^2} \Big) \log (z-w)  \, , \\
    X'^{\hat I}(z) Y(w, \bar w) \sim  & -A^{\hat I} y(z) y(w) \sim \frac{A^{\hat I}}{2R^2} \log (z-w) \, ,
\end{align}
as well as \eqref{eqn:XX} and \eqref{eqn:yyope} which remain unchanged. It is clear that for $R \to \infty$ the fields $X'^{\hat I}$ and $X^{\hat I}$ satisfy the same OPEs, and so the cases $A^{\hat I} = 0$ and constant $A^{\hat I} \ne 0$ are equivalent in the limit when the structure of the algebra is concerned, namely the currents \eqref{eqn:current} and \eqref{eqn:newcurrent} satisfy the same relations. 

For simplicity we will restrict to the case $A^{\hat I}=0$, described by \eqref{eqn:current} and \eqref{eqn:j}.
One can compute the OPE between two of such affine currents as follows.
Taking two states in the tower of the $E_8 \times E_8$ (or $SO(32)$) Cartan vectors
\begin{equation*}
\begin{split}
     J^{\hat{I}}_{n}(z) J^{\hat{J}}_{m}(w)  &= - :\partial X^{\hat I}(z) \, e^{i n y(z)}: \, :\partial X^{\hat J}(w)\, e^{i m y(w)}: \,   \\
    &= (z-w)^{\frac{ nm}{2 R^2}} :e^{in y(z)} e^{i m y(w)}  \Big( - \partial X^{\hat{I}}(z) \partial X^{\hat{J}}(w) + \frac{\delta^{\hat{I} \hat J}}{(z-w)^2} \Big): \, ,
    \end{split}
\end{equation*}
and expanding around $z=w$
\begin{equation}
\begin{split}
 &J^{\hat{I}}_{n}(z) J^{\hat{J}}_{m}(w) = \\
    &= (z-w)^{\frac{ nm}{2 R^2}} :\Big( 1 + i n \partial y(w) (z-w) + \ldots \Big) e^{i (n+m) y(w)}\Big( - \partial X^{\hat{I}}(z) \partial X^{\hat{J}}(w) + \frac{\delta^{\hat{I}\hat{J}}}{(z-w)^2} \Big): \,  \\
\label{eqn:survive}
    &\sim (z-w)^{\frac{ nm}{2 R^2}} \Big( \delta^{\hat{I}\hat J} \frac{:e^{i(n+m)y(w)}:}{(z-w)^2} + i \delta^{\hat I \hat J} n \frac{:\partial y(w) e^{i (n+m)y(w)}: }{z-w} + \mathcal{O}(1) \Big) \, ,
    \end{split}
\end{equation}
which is indeed \eqref{eqn:cc}. 

As for the OPE between two affine root currents
\begin{equation}
\label{eqn:a8}
\begin{split}
    J^{\alpha}_{n}(z) J^{\beta}_{m}(w) = &  :c_{\alpha} e^{i \pi_{\alpha}^{\hat I} X^{\hat I }(z)} e^{i n y(z)}: \, :c_{\beta} e^{i \pi_{\beta}^{\hat J} X^{\hat J}(w)} e^{i m y(w)}:  \\
    =& \,  c_{\alpha} c_{\beta}(z-w)^{ \pi_{\alpha} \cdot \pi_{\beta} + \frac{nm}{2R^2}} :e^{i \pi_{\alpha}^{\hat I} X^{\hat I}(z)} e^{i n y(z)}e^{i \pi_{\beta}^{\hat J} X^{\hat J}(w)} e^{i m y(w)}: \\
    = & \, c_{\alpha} c_{\beta}(z-w)^{ \pi_{\alpha} \cdot \pi_{\beta} + \frac{nm}{2R^2}} :\big[ 1+i \pi_{\alpha}^{\hat I} \partial X^{\hat I}(w) (z-w)+...)\cdot \\
    & \qquad \qquad \qquad \cdot (1+i n \partial y(w) (z-w)+\ldots \big]    e^{i(\pi_{\alpha}^{\hat I} + \pi_{\beta}^{\hat I})X^{\hat I}(w)} e^{i (n+m) y(w)}: \, .
\end{split}
\end{equation}
The explicit result depend on the pair of roots we consider. If $\pi_{\alpha}+\pi_{\beta}=\pi_{\alpha+\beta}$ still belongs to the root system of $E_8 \times E_8$ (or $SO(32)$), with the root normalization $|\pi|^2=2$ it holds $\pi_{\alpha} \cdot \pi_{\beta} = -1$ and \eqref{eqn:a8} reads
\begin{equation}
\begin{split}
        J^{\alpha}_n(z) J^{\beta}_m(w) \sim & (z-w)^{  \frac{nm}{2R^2}} \Big(   \frac{\epsilon(\alpha, \beta) c_{\alpha+\beta}:e^{i \pi_{\alpha+\beta}^{\hat I} X^{\hat I}(w)} e^{i (n+m) y(w)} :}{z-w} + \mathcal{O}(1) \Big)  \\
        = & (z-w)^{  \frac{nm}{2R^2}} \Big(\frac{\epsilon(\alpha, \beta) J^{\alpha+\beta}_{n+m}(w)}{z-w} + \mathcal{O}(1) \Big) \, .
\end{split}
\end{equation}
In the case $\pi_{\alpha}=-\pi_{\beta}$, $\pi_{\alpha} \cdot \pi_{\beta} = - |\pi_{\alpha}|^2=-2$, \eqref{eqn:a8} reads
\begin{equation}
\begin{split}
    J^{\alpha}_{n}&(z) J^{-\alpha}_{m}(w) \sim (z-w)^{  \frac{n m}{2R^2}} :e^{i (n+m) y(w)} \Big( \frac{1}{(z-w)^2} + \frac{i( \pi^{\hat I}_{\alpha} \partial X^{\hat I}(w) + n \partial y(w)) }{z-w} + \mathcal{O}(1) \Big): \, ,
\end{split}
\end{equation}
which can be rewritten as
\begin{equation}
    J^{\alpha}_n(z) J^{-\alpha}_m(w) \sim (z-w)^{  \frac{n m}{2R^2}} \Big( \frac{:e^{i (n+m) y(w)}:}{(z-w)^2} + \frac{\pi^{\hat I}_{\alpha} J^{\hat I}_{n+m}(w) + i n :\partial y(w)e^{i (n+m) y(w)}:}{z-w} + \mathcal{O}(1) \Big) \, .
\end{equation}
In all the other cases we do not find any integer pole for $R \to \infty$. These results can be summarized as
\begin{equation}
\label{eqn:rrgeneral}
    J^{\alpha}_{n}(z) J^{\beta}_{m}(w) \sim (z-w)^{\frac{nm}{2 R^2}} \cdot
    \begin{cases}
        \frac{\epsilon(\alpha, \beta) J^{\alpha+\beta}_{n+m}(w)}{z-w} + \mathcal{O}(1)\qquad \qquad \qquad \qquad \qquad \qquad \quad \, \,   \alpha+\beta \,  {\rm root,} \\
        \frac{:e^{i (n+m) y(w)}:}{(z-w)^2} + \frac{\pi_{\alpha}^{\hat{I}} J^{\hat{I}}_{n+m}(w)+in:\partial y(w) e^{i (n+m)y(w)}:}{z-w} +\mathcal{O}(1) \, \, \, \, \alpha=-\beta\, , \\
        \mathcal{O}(1) \qquad \qquad \qquad \qquad \qquad \qquad \qquad\qquad \qquad \qquad {\rm \, otherwise,} 
    \end{cases}
\end{equation}
which is \eqref{eqn:rr}. 

For the OPE between one current in the Cartan tower and the other in the root tower
\begin{equation}
\begin{split}
    J^{\hat I}_n(z) J^{\alpha}_{m}(w)= & :i \partial X^{\hat I}(z) e^{i n y(z)}: \, :c_{\alpha} e^{i  \pi_{\alpha}^{\hat J} X^{\hat J}(w)} e^{i m y(w)}: \\
    = & \, i c_{\alpha} (z-w)^{\frac{nm}{2 R^2}}  \frac{ -i \pi^{\hat I}_{\alpha} :e^{i \pi^{\hat J}_{\alpha} X^{\hat J}(w)} e^{in y(z)} e^{i m y(w)}:}{z-w}  \\
    \sim & \,  c_{\alpha} (z-w)^{\frac{nm}{2 R^2}}  \frac{\pi^{\hat I}_{\alpha} :e^{i \pi^{\hat J}_{\alpha} X^{\hat J}(w)} e^{i (n+m) y(w)}: }{z-w} = \frac{\pi_{\alpha}^{\hat I} J^{\alpha}_{n+m}(w)}{z-w}  \, ,
    \end{split}
\end{equation}
as $R \to \infty$. Again, this is consistent with \eqref{eqn:cr}. \newline
\eqref{JIdy}, \eqref{Jalphady} and \eqref{dydy} are straightforward from \eqref{eqn:yyope}.

From the OPEs between the currents one can compute the algebra between their zero modes \eqref{eqn:generators} as follows. The commutator between two Cartan generators for $R \to \infty$ reads
\begin{equation}
    [(J^{\hat I}_n)_0, (J^{\hat J}_m)_0] = \oint_{C'} \frac{dz}{2\pi i} \oint_C \frac{dw}{2\pi i}J^{\hat I}_n(z)  J^{\hat J}_m(w) - \oint_{C'} \frac{dw}{2\pi i} \oint_C \frac{dz}{2\pi i} J^{\hat I}_n(z) J^{\hat J}_m(w)\, 
\end{equation}
where $C'$ is a contour external to $C$. Then one obtains
\begin{equation}
\label{eqn:cases}
\begin{split}
    [(J^{\hat I}_n)_0, (J^{\hat J}_m)_0]  = & \oint \frac{dw}{2\pi i} \delta^{\hat I \hat J} \mathrm{Res}_{z \to w} \Big[ \frac{:e^{i(n+m)y(w)}:}{(z-w)^2} + i n \frac{:\partial y(w) e^{i (n+m) y(w)}: }{z-w} \Big] \\
    = & \oint \frac{dw}{2\pi i } in \delta^{\hat I \hat J} \partial y(w) e^{i (n+m) y(w)} \\
     = &
    \begin{cases}
        \oint \frac{dw}{2\pi i} i n \delta^{\hat I \hat J} \partial y(w) = in \delta^{\hat I \hat J} (\partial y)_0 \text{ for } n+m=0 \, ,\\
         \oint \frac{dw}{2 \pi i } \delta^{\hat I \hat J} \frac{n}{n+m} \partial(e^{i(n+m)y(w)})=0 \text{ for } n+m\ne 0 \, 
    \end{cases} \\
    = & \,  i n \delta^{\hat I \hat J} \delta_{n+m, 0} (\partial y)_0 \, .
    \end{split}
\end{equation}
The commutator for a Cartan generator and a ladder operator is
\begin{equation}
\begin{split}
    [(J^{\hat I}_n)_0, (J^{\alpha}_{m})_0] = & \oint_{C'} \frac{dz}{2\pi i} \oint_C \frac{dw}{2\pi i}J^{\hat I}_n(z)  J^{\alpha}_{m}(w) - \oint_{C'} \frac{dw}{2\pi i} \oint_C \frac{dz}{2\pi i} J^{\hat I}_n(z) J^{\alpha}_{m}(w) \\
    = & \oint \frac{dw}{2\pi i} \mathrm{Res}_{z \to w} \Big[ \frac{\pi_{\alpha}^{\hat I} c_{\alpha} :e^{i \pi^{\hat J}_{\alpha}X^{\hat J}(w)} e^{i  (n+m) y(w)}:}{z-w} \Big] \\
    = &  \oint \frac{dw}{2\pi i} \pi_{\alpha}^{\hat I} c_{\alpha} :e^{i \pi^{\hat J}_{\alpha}X^{\hat J}(w)} e^{i  (n+m) y(w)}: \\
    =& \pi_{\alpha}^{\hat I} (J^{\alpha}_{n+m})_0 \, , 
    \end{split}
\end{equation}
and finally the commutator between two ladder operators is
\begin{equation}
\begin{split}
   & [(J^{\alpha}_{n})_0, (J^{\beta}_{m})_0] = \oint_{C'} \frac{dz}{2\pi i} \oint_C \frac{dw}{2\pi i}J^{\alpha}_{n}(z)  J^{\beta}_{m}(w) - \oint_{C'} \frac{dw}{2\pi i} \oint_C \frac{dz}{2\pi i} J^{\alpha}_{n}(z)  J^{\beta}_{m}(w)   \\
    & \ = \oint \frac{dw}{2\pi i} \mathrm{Res}_{z \to w} 
    \begin{cases}
          \frac{\epsilon(\alpha, \beta) c_{\alpha + \beta} e^{i (\pi^{\hat I}_{\alpha +\beta}) X^{\hat I}(w)}e^{i(n+m)y(w)}}{z-w} + \mathcal{O}(1)\qquad \qquad \qquad \qquad \quad \, \,\,\, \, \alpha+\beta \,  {\rm root,} \\
        \frac{:e^{i (n+m) y(w)}:}{(z-w)^2} + \frac{:i \pi^{\hat I}_{\alpha} \partial X^{\hat I}(w) e^{i (n+m) y(w)} + i n \partial y(w) e^{i (n+m) y(w)}:}{z-w} +\mathcal{O}(1) \, \, \, \, \, \alpha=-\beta, \\
        \mathcal{O}(1) \qquad \qquad \qquad \qquad \qquad \qquad  \qquad\qquad \qquad \qquad \qquad \qquad \, \, {\rm otherwise.} 
    \end{cases}\\
    &\qquad \qquad  \qquad \  = \begin{cases}
          \epsilon(\alpha, \beta) (J^{\alpha + \beta}_{n+m})_0  \qquad  \qquad \quad \quad  \,\,  \,  \alpha+\beta \, \, \,  {\rm root} \\
          \pi^{\hat I}_{\alpha} (J^{\hat I}_{n+m})_0 + i n \delta_{n+m, 0} (\partial y)_0 \quad \alpha=-\beta\\
         0  \qquad \qquad \qquad \qquad \qquad \qquad \, \, \, \, {\rm otherwise.} 
    \end{cases}
    \end{split}
\end{equation}

\section{Test of the RFC by dimensional reduction}
\label{app:RFC}
In this appendix we review the dimensional reduction of the 10d heterotic supergravity Lagrangian in detail, as well as the calculation of the long-range interactions between heterotic states in the field theory language.

\subsection{Dimensional reduction of the supergravity Lagrangian}
We consider an $S^1$ compactification of the heterotic string from 10 to 9 dimensions, by identifying the 9th coordinate as $x^9 \equiv y \sim y+ 2 \pi R$. From the target spacetime point of view, the low energy heterotic supergravity action in 10 dimensions takes the form \cite{homh} 
\begin{equation}
\label{eqn:16}
\begin{split}
    S= \frac{1}{(2\pi)^7 \alpha'^{4}} \int d^{10}x \sqrt{-\boldsymbol{G}^s} e^{-2 \boldsymbol{\Phi}} \Big(\boldsymbol{R}+4\partial_{M}\boldsymbol{\Phi} \partial^M \boldsymbol{\Phi} -\frac{1}{12} \boldsymbol{H}^{MNR} \boldsymbol{H}_{MNR}-\\
    - \frac{\alpha'}{4}  \boldsymbol{\bar{F}}^{MN}_I  \boldsymbol{\bar{F}}_{MN}^I \Big) \, ,
    \end{split}
\end{equation}
$\boldsymbol{G}^s _{MN}$, $M,N=0,...,9$ is the 10 dimensional metric in the string frame with Ricci scalar $\boldsymbol{R}$, $\boldsymbol{\Phi}= \tilde{\boldsymbol{\Phi}} + \Phi_0 $ is the 10 dimensional dilaton and the gauge invariant field strengths are defined as
\begin{align}
\label{eqn:18}
     \boldsymbol{H}_{MNR} &= 3 \Big( \partial_{[M} \boldsymbol{B}_{NR]}- \alpha' \boldsymbol{A}^{I}_{[M}\partial_N  \boldsymbol{A}_{R]I}-\frac{\alpha'}{3} f_{IJK}  \boldsymbol{A}^{I}_{M}  \boldsymbol{A}^{J}_{N}  \boldsymbol{A}^{K}_{R}\Big)\, ,\\
\label{eqn:19}
     \boldsymbol{F}_{MN}^{I} & = \sqrt{\alpha'} (2 \partial_{[M}  \boldsymbol{A}^{I}_{N]} + f_{JK}^{I}  \boldsymbol{A}^{J}_{M}  \boldsymbol{A}^{K}_{N}) = \sqrt{\alpha'} \boldsymbol{\bar{F}}_{MN}^{I} \, ,
\end{align}
where $\boldsymbol{B}_{MN}$ is the NSNS 2-form and $\boldsymbol{A}^I$ are the heterotic vector bosons, with $I=1,...,\, 496$ the gauge group index (taking into account both the Cartan and the root sectors). 
For the reduction, the spacetime indices are split as $M=(\mu, 9)$,  with $\mu=0,...,8$ and we take the following ansatz for the metric (in the string frame) 
\begin{equation}
\label{eqn:17}
    ds^2= \boldsymbol{G^s}_{MN} dx^{M} dx^N=e^{\frac{4}{7} \tilde \Phi} e^{-\frac{2}{7}\tilde \sigma}g_{\mu \nu} dx^{\mu} dx^{\nu} + e^{2 \tilde \sigma} (dy + Z_{\mu} dx^{\mu})^2 \, .
\end{equation}
Here $g_{\mu \nu}$ is the 9 dimensional metric in the Einstein frame, $Z_{\mu}$ the (dimensionless) graviphoton, and $\tilde \sigma$ is the dynamical part of the (dimensionless) radion field $\sigma = \tilde \sigma + \sigma_0$, where $\sigma_0$ is the background value. The physical compactification radius of the circle is therefore $\mathcal{R}=e^{\tilde \sigma}R$.  

We restrict to the massless sector upon compactification and therefore assume the 10 dimensional fields (in bold) to be independent of the coordinate $y$. The dimensionally reduced action in the Einstein frame in terms of the 9 dimensional fields (not in bold) reads 
\begin{equation}
\label{eqn:24}
\begin{split}
    S=&\frac{e^{-2 \Phi_0} R}{(2\pi)^6 \alpha^{\prime 4}}\int d^{9}x \sqrt{-g} \Big( R- \frac{8}{7} \partial_{\mu} \tilde{\sigma} \partial^{\mu} \tilde{\sigma} - \frac{4}{7} \partial_{\mu} \tilde{\Phi} \partial^{\mu} \tilde{\Phi} + \frac{4}{7} \partial_{\mu} \tilde{\Phi} \partial^{\mu} \tilde{\sigma} -   \\      
     &-\frac{\alpha'}{2} e^{- 2 \tilde \sigma} \bar{F}_{\mu 9}^I \bar{F}^{\mu I}_9- \frac{1}{4} e^{\frac{16}{7} \tilde{\sigma}} e^{-\frac{4}{7} \tilde{\Phi}} Z_{\mu \nu} Z^{\mu \nu} - \frac{\alpha'}{4} e^{-\frac{4}{7} \tilde{\Phi}} e^{\frac{2 }{7}\tilde{\sigma}} \bar{F}_{\mu \nu}^I \bar{F}^{\mu \nu I}  -\\
     &-\frac{1}{12} e^{\frac{4 }{7}\tilde{\sigma}} e^{-\frac{8}{7}\tilde{\Phi}} H_{\mu \nu \rho} H^{\mu \nu \rho} -\frac{1}{4} e^{-\frac{12 }{7}\tilde{\sigma}} e^{-\frac{4}{7}\tilde{\Phi}} H_{\mu \nu 9} H^{\mu \nu }_9 \Big),
     \end{split}
\end{equation}
 where
\begin{align}
\label{eqn:20}
    \bar{F}_{\mu 9}^{I} &= \partial_{\mu}  A^{I}_{9} + f_{JK}^{I}  A^{J}_{\mu}  A^{K}_{9}\, , \\
\label{eqn:21}
    \bar{F}_{\mu \nu}^{I} &= 2 \partial_{[\mu} A_{\nu]}^I + f_{JK}^{I}  A^{J}_{\mu}  A^{K}_{\nu} + Z_{\mu \nu} A_{9}^I \, , 
\end{align}
and $Z_{\mu \nu}= \partial_{\mu}Z_{\nu}-\partial_{\nu}Z_{\mu}$. By defining also
\begin{equation}
    W_{\mu}=\boldsymbol{B}_{\mu 9} + \frac{\alpha'}{2} A_9^I A_{\mu I}\, , \qquad W_{\mu \nu} = \partial_{\mu} W_{\nu} - \partial_{\nu} W_{\mu} \, ,
\end{equation}
we obtain
\begin{align}
\label{eqn:22}
    H_{\mu \nu 9} & = W_{\mu \nu} - \alpha' (2 \partial_{[\mu} A_{\nu]}^I + f_{JK}^{I}  A^{J}_{\mu}  A^{K}_{\nu}) A_{9I} - \frac{\alpha'}{2} Z_{\mu \nu} (A_{9}^I)^2 \, , \\
H_{\mu \nu \rho}& = 3 \Big(\partial_{[\mu} B_{\nu \rho]} - \alpha'A^I_{[\mu}\partial_{\nu}A_{\rho]I}- (\partial_{[\mu}Z_{\nu})W_{\rho]}+ Z_{[\mu} (\partial_{\nu}W_{\rho]})-\frac{\alpha'}{3}f_{IJK}A_{\mu}^IA_{\nu}^{J}A_{\rho}^K \Big)\, . 
\end{align}
In the following, we choose to work with dimension 1 gauge fields, normalized so that string states have integer quantized charges. The momentum number is derived by looking at the diffeomorphism symmetry of the 10 dimensional action. If one wants the charge of the $n$-th KK state from the expansion
\begin{equation}
    \phi\left( x^M \right)\, =\,  \sum_n \phi_n(x^{\mu}) e^{i\frac{n y}{R}}
\end{equation}
to be $n \in \mathbb{Z}$ under the redefined field $\tilde{Z}_{\mu}$, the dimensionfull graviphoton must be taken as
\begin{equation}
\label{eqn:26}
    \tilde{Z}_{\mu} = \frac{Z_{\mu}}{R} \, .
\end{equation}
 The expressions for the 9 dimensional heterotic field strength \eqref{eqn:21} is thus
\begin{equation}
\label{eqn:27}
    \bar{F}_{\mu \nu}^{I}=  2 \partial_{[\mu} A_{\nu]}^I + 2 R \partial_{[\mu} \tilde{Z}_{\nu]} A_{9}^I + f_{JK}^{I}  A^{J}_{\mu}  A^{K}_{\nu}\, .
\end{equation}
Focusing now on the $W_{\mu}$ redefinition and the winding charge, its expression can be derived from the worldsheet action term \cite{palti}
\begin{equation}
    S=-\frac{1}{2\pi \alpha'} \int d\tau d\Sigma \,  W_{\mu} \partial_{\tau} X^{\mu} \partial_{\Sigma}Y\, ,
\end{equation}
where $\left( \tau, \, \Sigma \right)$ are the coordinates parameterizing the worldsheet, $\Sigma \in [0, 2 \pi)$, and $X^M \left( \tau, \, \Sigma \right)$ are the embedding functions into target spacetime.
Since the string winds around the $S^1$ direction $w$ times, one can write
\begin{equation}
    Y=w\Sigma R \, .
\end{equation}
The worldsheet action then yields
\begin{equation}
    S= -\frac{w R}{\alpha'} \int d\tau W_{\mu} \partial_{\tau}X^{\mu} \, ,
\end{equation}
which is the worldline coupling of a particle with charge 
\begin{equation}
    Q = \frac{w R}{\alpha'}
\end{equation}
with the dimensionless gauge field $W_{\mu}$. To get an integer charge, the dimensionful gauge field $\tilde{W}_{\mu}$ must be defined as
\begin{equation}
\label{eqn:28}
   \tilde{W}_{\mu} = \frac{W_{\mu}R}{\alpha'}\, . 
\end{equation}
With the redefinitions \eqref{eqn:26} and \eqref{eqn:28}, the 9 dimensional field strengths obtained by reducing $\boldsymbol{H}_{MNR}$ are
\begin{equation}
\begin{split}
    H_{\mu \nu 9} =\frac{\alpha'}{R} (\partial_{\mu} \tilde{W}_{\nu} - \partial_{\nu} \tilde{W}_{\mu} ) - \alpha'(\partial_{\mu} A_{\nu}^I)A_{9}^I + \alpha'(\partial_{\nu} A_{\mu}^I)A_{9}^I  - \frac{\alpha' R}{2} \partial_{\mu}\tilde{Z}_{\nu} (A_{9}^I)^2  + \\
 + \frac{\alpha' R}{2} \partial_{\nu} \tilde{Z}_{\mu} (A_{9}^I)^2 - \alpha' f_{IJK} A_{\mu}^I A_{\nu}^J A_{9}^K   
\end{split}
\end{equation}
and
\begin{equation}
    H_{\mu \nu \rho} = 3 \Big(\partial_{[\mu} B_{\nu \rho]} - \alpha'A^I_{[\mu}\partial_{\nu}A_{\rho]}^I-\alpha' ((\partial_{[\mu}\tilde{Z}_{\nu})\tilde{W}_{\rho]}+ \tilde{Z}_{[\mu} (\partial_{\nu}\tilde{W}_{\rho]})) -\frac{\alpha'}{3}f_{IJK}A_{\mu}^IA_{\nu}^{J}A_{\rho}^K \Big). 
\end{equation} 
Furthermore, the Cartan heterotic gauge fields $A^{\hat{I}}$ are correctly normalized in such a way for the charges to be $\pi^{\hat{I}} \in \Gamma_8 \times \Gamma_8$. Finally, in order to work with conventionally normalized scalar fields, without cross terms in the kinetic part, we make the following field redefinition
\begin{equation}
\label{eqn:radionred}
    \lambda = \frac{4}{\sqrt{7}}\Big(\frac{\Phi}{4} - \sigma \Big) \longrightarrow \lambda_0 = \frac{4}{\sqrt{7}}\Big(\frac{\Phi_0}{4} - \sigma_0 \Big) \, , \, \tilde \lambda = \frac{4}{\sqrt{7}}\Big(\frac{\tilde \Phi}{4} - \tilde \sigma \Big) \, .
\end{equation}
Finally, the 9 dimensional action in the Einstein frame, with conventionally normalized scalars and gauge fields such that minimally coupled sources have integer charges takes the form 
\begin{equation}
\label{eqn:9daction}
\begin{split}
    S=&\dfrac{\Mp9^{7}}{2}  \int d^{9}x \sqrt{-g} \Big( R- \frac{1}{2} \partial_{\mu} \lambda \partial^{\mu} \lambda - \frac{1}{2} \partial_{\mu} \Phi \partial^{\mu} \Phi - \frac{\alpha'}{2} e^{- \frac{\tilde \Phi}{2}} e^{\frac{\sqrt{7}}{2}\tilde \lambda} \bar{F}_{\mu 9}^I \bar{F}^{\mu I}_9- \frac{1}{4} e^{-\frac{4\sqrt{7}}{7} \tilde \lambda} R^2 Z_{\mu \nu} Z^{\mu \nu}- \\
     & - \frac{\alpha'}{4} e^{-\frac{1}{2} \tilde \Phi} e^{-\frac{\sqrt{7}}{14}\tilde \lambda} \bar{F}_{\mu \nu}^I \bar{F}^{\mu \nu I}  -\frac{1}{12} e^{-\frac{\sqrt{7}}{7} \tilde \lambda} e^{- \tilde \Phi} H_{\mu \nu \rho} H^{\mu \nu \rho} -\frac{1}{4} e^{\frac{3\sqrt{7}}{7} \tilde \lambda} e^{-\tilde \Phi} H_{\mu \nu 9} H^{\mu \nu}_9 \Big),
     \end{split}
\end{equation}
where the 9 dimensional Planck mass reads
\begin{equation}
\dfrac{\Mp9^{7}}{2}=\frac{e^{-2\Phi_0} R}{(2\pi)^6 \alpha^{\prime 4}}=\frac{e^{-\frac{7}{4}(\Phi_0+\frac{1}{\sqrt{7}}\lambda_0)} }{(2\pi)^6 \alpha^{\prime 4}} \, .
\end{equation} 

To close this section, let us recast the mass formula for the heterotic string states in the 9 dimensional Einstein frame and in Planck units and in the absence of Wilson lines (with $R= e^{\sigma_0} \sqrt{\alpha'}$)  
\begin{equation}
\label{eqn:mpunits}
    \dfrac{M^2}{\Mp9^2}= (32\pi^6)^{\frac{2}{7}} \left\{ e^{\frac{4\sqrt{7}}{7}\lambda} n^2+ e^{\Phi} e^{-\frac{3\sqrt{7}}{7}\lambda} w^2  + e^{\frac{1}{2} \Phi} e^{\frac{\sqrt{7}}{14} \lambda} \left[ 2\left(N+\bar{N} -\frac{3}{2}\right) + |\pi|^2 \right] \right\}.
\end{equation}
We also introduce the following notation for shortness \begin{equation}
    M^2 = M^2_n + M^2_w+M^2_N+M^2_{\pi} \, ,
\end{equation}
where
\begin{align}
\label{eqn:n}
    M^2_n & = (32\pi^6)^{\frac{2}{7}}  e^{\frac{4 \sqrt{7}}{7} \lambda} n^2 \Mp9^2\, ,\\
    M^2_w & = (32\pi^6 )^{\frac{2}{7}} e^{\Phi} e^{-\frac{3 \sqrt{7}}{7}\lambda}w^2 \Mp9^2\, ,\\
    M^2_N & = 2 (32\pi^6 )^{\frac{2}{7}} e^{\frac{1}{2}\Phi} e^{\frac{\sqrt{7}}{14}\lambda}  \Big(N+\bar{N} -\frac{3}{2}\Big) \Mp9^2\, , \\
\label{eqn:pi}
    M^2_{\pi} & = (32\pi^6 )^{\frac{2}{7}} e^{\frac{1}{2}\Phi} e^{\frac{\sqrt{7}}{14}\lambda}  |\pi|^2 \Mp9^2\, .
\end{align}

\subsection{Computing the long range force}
In order to test the RFC in this setting, we need to compute the long range force between two states in the heterotic string spectrum as prescribed in e.g. \cite{force}. The part coming from the $U(1)$ interactions, in the system at hand and with the conventions introduced above, is mediated by $\tilde{Z}_{\mu}$, $\tilde{W_{\mu}}$ and the Cartan vectors of $E_8 \times E_8$, $A^{\hat{I}}$, $\hat{I}=1,...,16$.
This perturbative computation makes sense only when the theory is weakly coupled, and we have just argued that in the two corners of interest in moduli space actually there is at least one divergent gauge coupling. This means that we cannot trust the full computation at all points in moduli space if the three interactions are turned on at the same time. Nevertheless, we will perform the computation including all gauge fields at the same time for completeness because it is valid as an abstract calculation, but we will then limit ourselves to its application to states charged only under the groups whose gauge coupling is perturbative in the region of moduli space that we want to study.

Consider then a generic state, with mass given by eq.\eqref{eqn:mpunits}, minimally coupled  to gravity and to the gauge fields. The leading contribution to the force is obtained through the linearization of action \eqref{eqn:9daction}, which yields
\[
    S=\frac{\Mp9^7}{2} \int d^{9}x \sqrt{-g} \Big( R- \frac{1}{2} \partial_{\mu} \lambda \partial^{\mu} \lambda - \frac{1}{2} \partial_{\mu} \Phi \partial^{\mu} \Phi  -\frac{\alpha'}{2}  \partial_{\mu} A_{9}^{\hat{I}} \partial^{\mu} A^{\hat{I}}_9 - \frac{R^2}{4} \tilde{Z}_{\mu \nu} \tilde{Z}^{\mu \nu} -       
\]
\begin{equation}
\label{eqn:90}
     -  \frac{\alpha'}{4} \tilde{F}_{\mu \nu}^{\hat{I}} \tilde{F}^{\mu \nu \hat{I}}  -\frac{\alpha'^2}{4R^2} \tilde{W}_{\mu \nu} \tilde{W}^{\mu \nu} \Big) - \int M(\Phi, \lambda, A_9^I) \, ds - w \int \tilde{W} - \pi^{\hat{I}} \int A^{\hat{I}} - n \int \tilde{Z},
\end{equation}
where 
\begin{equation}
    \tilde{F}_{\mu \nu}^{\hat{I}} = 2 \partial_{[\mu} A_{\nu]}^{\hat{I}} \, .
\end{equation}

We expand the Einstein frame metric around the Minkowski background, that is
\begin{equation}
    g_{\mu \nu} = \eta_{\mu \nu} + h_{\mu \nu}\, ,
\end{equation}
where $h_{\mu \nu} << 1$ in Planck units. We will use the trace-reversed metric perturbation
\begin{equation}
\label{eqn:31}
    \bar{h}_{\mu \nu} = h_{\mu \nu} - \frac{1}{2} \eta_{\mu \nu} h \, ,
\end{equation}
where $h\equiv \eta^{\mu \nu} h_{\mu \nu}$, and under the hypothesis of a static solution, so that $\partial_t$ is a Killing vector defining a conserved energy. 
Furthermore, we work in the Lorentz gauge for all the gauge fields and also for the trace-reversed metric perturbation: $\partial_{\mu} \bar{h}^{\mu \nu} =0$,   $\partial_{\mu} \tilde{W}^{\mu \nu} =0$, $\partial_{\mu} \tilde{Z}^{\mu \nu} =0$, $\partial_{\mu} \tilde F^{\mu \nu I} =0$. 

We now compute the perturbation on the background caused by a source particle in the heterotic string spectrum at rest,  with worldline parameterized by its proper time as
\begin{equation}
    x^{\mu}(\tau) = (\tau, \hat{x}^i)\, ,  \qquad \hat{x}^i=\mathrm{const.}
\end{equation}
The linerarized equations of motion and their respective solutions thus read (where $M \equiv M(\Phi_0, \lambda_0,0$)  and also all the derivatives of the mass are evaluated at the background)
\begin{align}
\label{eqn:peq1}
    \Box \bar{h}_{\mu \nu} (x) = - \frac{2\, M}{\Mp9^7}\,   \,\delta_{\mu}^0\, \delta_{\nu}^0 \, \delta^{(8)}(\hat{x}^i-x^i) &\implies \bar{h}_{\mu \nu}(r)=\dfrac{M}{3 \, V_7\, \Mp9^7\, r^6} \,\delta^0_{\mu} \, \delta^0_{\nu} \, ,\\
    \Box \lambda(x) =  \frac{2}{\Mp9^7} \frac{\partial M}{\partial \lambda} \, \delta^{(8)}(\hat{x}^i-x^i) & \implies \tilde \lambda(r)= -\dfrac{ \partial_{\lambda} M}{3\, V_7\, \Mp9^7\, r^6} \, ,\\
    \Box A_9^{\hat{I}}(x)= \frac{2}{\Mp9^7} \, \frac{1}{\alpha'}\,  \frac{\partial M}{\partial A_{\hat{I}}^9}\, \delta^{(8)}(\hat{x}^i-x^i) & \implies A_9^{\hat{I}}(r)=-\dfrac{\partial_{A_{\hat{I}}^9} M}{3 \, \alpha' \, V_7\, \Mp9^7\, r^6} \, , \\
    \Box \Phi(x) = \frac{2}{\Mp9^7} \frac{\partial M}{\partial \Phi} \,  \delta^{(8)}(\hat{x}^i-x^i) & \implies \tilde{\Phi}(r)=-\dfrac{ \partial_{\Phi} M}{3\, V_7\, \Mp9^7\, r^6} \, , \\
    \Box A^{\mu \hat{I}}(x) = \frac{2}{\Mp9^7}\,  \frac{1}{\alpha'}\,  \pi^{\hat{I}} \, \delta^{\mu}_0 \, \delta^{(8)}(\hat{x}^i-x^i) & \implies A^{\mu \hat{I}}(r)=-\dfrac{\pi^{\hat{I}}}{3\, \alpha'\, V_7\, \Mp9^7\, r^6}\, \delta^{\mu}_0 \, ,\\
    \Box \tilde{W}^{\mu}(x) = \frac{2}{\Mp9^7} \,  \frac{R^2}{\alpha'^2} \, w \, \delta^{\mu}_0  \delta^{(8)}(\hat{x}^i-x^i) & \implies \tilde{W}^{\mu}(r)=-\dfrac{w\, R^2}{3\, \alpha'^2\, V_7 \, \Mp9^7\, r^6}\delta^{\mu}_0 \, ,\\
\label{eqn:peq2}
    \Box \tilde{Z}^{\mu}(x) = \frac{2}{\Mp9^7}\,  \frac{1}{R^2}\,  n \,  \delta^{\mu}_0 \,  \delta^{(8)}(\hat{x}^i-x^i) & \implies \tilde{Z}^{\mu}(r)=-\dfrac{n}{3\, R^2\, V_7\, \Mp9^7\, r^6}\, \delta^{\mu}_0 \,.  
\end{align}
These solutions are taken to be static ($\Box = \vec{\nabla}^2$) and such that all the field perturbations vanish at infinity. $V_7$ is the volume of the $7$ dimensional unit sphere, given by
\begin{equation}
    V_{7}=\frac{2\pi^{4}}{\Gamma(4)} \, ,
\end{equation}
and we defined the radial coordinate $r=|x^i- \hat{x}^i|$. 

The force felt by a probe particle with mass $M_2$ and charges $(n_2, \, w_2, \, \pi^I_2)$  sitting a large distance $r$ away from another particle with mass $M_1$ and charges $(n_1, \, w_1, \, \pi^I_1)$, can be derived from the potential felt by the probe
\begin{equation}
    V_{12}=\left( 1-\frac{h_{00}(r)}{2}\right) M_2(\Phi(r),\lambda(r), A_9^{\hat{I}}(r))+w\, \tilde{W}_0(r)+\pi^{\hat{I}} A_0^{\hat{I}}(r)+ n\, \tilde{Z}_0(r) \, ,
\end{equation}
where from the definition \eqref{eqn:31} we have
\begin{equation}
    h_{\mu \nu} = \bar{h}_{\mu \nu} -\frac{1}{7}g_{\mu \nu} \bar{h} \implies  h_{00} =\frac{6}{7} \bar{h}_{00}  \, .
\end{equation}
By expanding the mass around the background and keeping only the leading terms in $r$ one gets 
\begin{equation}
\begin{split}
    V_{12} =& - \dfrac{h_{00}(r)}{2} M_2(\Phi_0,\sigma_0,0) + \dfrac{\partial M_2}{\partial \Phi}(\Phi_0,\sigma_0,0) \tilde{\Phi}(r)+\dfrac{\partial M_2}{\partial \sigma}(\Phi_0,\sigma_0,0) \tilde{\sigma}(r)+ \\
    &+ \dfrac{\partial M_2}{\partial A_{9}^{\hat{I}}}(\Phi_0,\sigma_0,0) A_9^{\hat{I}}(r) - w \, \tilde{W}^0(r)-\pi^{\hat{I}} A^{0 \hat{I}}(r) - n\, \tilde{Z}^0(r) + \mathcal{O}(r^{-12}) \, .
\end{split}
\end{equation}
By defining $M_{i} \equiv M_i(\Phi_0, \lambda_0, 0)$ for $i=1$ and $2$, and evaluating all the derivatives of the mass at the background once again, the force takes the form
\begin{equation}
\label{eq:V12}
\begin{split}
     F_{12}=-\dfrac{\partial V_{12}}{\partial r} =& \dfrac{2}{V_7\, \Mp9^7\, r^7} \Bigg\{- \dfrac{3}{7}M_1\, M_2 - \dfrac{\partial M_1}{\partial \Phi} \, \dfrac{\partial M_2}{\partial \Phi} - \dfrac{\partial M_1}{\partial \lambda}\dfrac{\partial M_2}{\partial \lambda} - \frac{1}{\alpha'}  \dfrac{\partial M_1}{\partial A_{\hat{I}}^9}\dfrac{\partial M_2}{\partial A_{9}^{\hat{I}}} +\\
   &+ \dfrac{w_1 w_2 e^{\frac{1}{2}\Phi_0} e^{\frac{\sqrt{7}}{2}\lambda_0}}{\alpha'} +  \dfrac{\pi^{\hat{I}}_1 \pi^{\hat{I}}_2 }{\alpha'} + \dfrac{n_1 n_2 e^{-\frac{1}{2}\Phi_0} e^{-\frac{\sqrt{7}}{2}\lambda_0}}{\alpha'} \Bigg\}\, ,
   \end{split}
\end{equation}
and one can see that the scalar and gravitational forces give attractive contributions whereas the $U(1)$'s give repulsive ones, as expected for particles with equal charges. 

Keeping the 9 dimensional Planck mass fixed, the derivatives of the mass evaluated at the background read
\begin{align}
    \dfrac{\partial M }{\partial \Phi} &= \frac{1}{2 M} \left[ M^2_w + \frac{1}{2} M^2_{\pi}\right]\, ,\\
    \dfrac{\partial M}{\partial \lambda}& = \frac{1}{2 M} \left[ \frac{4 \sqrt{7}}{7} M^2_n-\frac{3 \sqrt{7}}{7}M^2_w+\frac{\sqrt{7}}{14} (M^2_N+M^2_{\pi}) \right]\, , \\
    \dfrac{\partial M}{\partial A^{\hat{I}}_9} &=\, \sqrt{\alpha'} \ \dfrac{ M_{\pi \hat{I}} (M_w - M_n)}{M} \, ,
\end{align}
so that, combining all the contributions 
\begin{equation}
\label{eqn:33}
\begin{split}
     V_7 \, \Mp9^7 \, r^7 \,  F_{12}(r)=& - M_1 M_2 +2M_{\pi,1}\cdot M_{\pi,2}+2M_{n,1}\cdot M_{n,2}+2M_{w,1}\cdot M_{w,2} -\\ 
     &-\dfrac{(M_{n,1}^2-M_{w,1}^2)(M_{n,2}^2-M_{w,2}^2)}{M_1 M_2}-\\ &-\dfrac{2M_{\pi,1}M_{\pi,2}(M_{n,1}-M_{w,1})(M_{n,2}-M_{w,2})}{M_1 M_2}\, .
    \end{split}
\end{equation}
There are several interesting cases in which this force vanishes, namely
\begin{itemize}
    \item[(i)]{For states having only momentum mass $M_{n}\ne 0$, that is $M_{i}=M_{n,i}$ the force takes the form $F_{12}\propto-M_{n,1} M_{n,2}-M_{n,1}M_{n,2}+2M_{n,1}M_{n,2}=0$. The same holds for states having only winding charge, $M_i=M_{w,i}$.}
    \item [(ii)]{The force between one momentum state and one winding state vanishes as well: $F_{12}\propto-M_{1,n}M_{2,w}+M_{1,n}M_{2,w}=0$.}
    \item[(iii)]{Since the states we are most interested in (namely the ones that form the \mbox{$(E_9 \oplus E_9)/\sim$} algebra) are BPS, let us consider the case of BPS heterotic states, which are characterized by the conditions \cite{4}
    \begin{equation}
    \label{eqn:bps}
        M^2= \frac{2}{\alpha'}p_R^2 \stackrel{\text{NS sector}}{\implies} \bar{N}=\frac{1}{2}, \quad N=1-nw-\frac{|\pi|^2}{2} \, .
    \end{equation}
    The force between two mutually BPS states vanishes, as can be easily seen by recasting the expression \eqref{eqn:33} in the equivalent form (also found in \cite{3})
    \begin{equation}
    V_7 r^7 F_{12} = - \dfrac{4}{\alpha'^2\,  M_1 \, M_2}\Big(\frac{\alpha'}{2} M_1 M_2 - \boldsymbol{p}_{L,1} \cdot \boldsymbol{p}_{L,2}\Big)\Big(\frac{\alpha'}{2} M_1 M_2 - p_{R,1} p_{R,2}\Big)\, .
\end{equation}
    From the second factor, mutually BPS particles exert a vanishing force on each other.}
\end{itemize}

The RFC states that for each point in the charge lattice there must exist a particle such that the force between two identical particles is non-negative, i.e. $F_{11} \geq 0$. From \eqref{eqn:33}, the state  having 
$N-\bar{N}-\frac{3}{2}=0$ is self repulsive and it is among the ones predicted by the RFC. Indeed, in this case 
\begin{equation}
    V_7 \Mp9^7 r^7 M^2 F_{11}(r) = (M_{\pi}^2 + 2 M_{w}M_n)^2 >0.
\end{equation}
Since the factor $V_7 \Mp9^7 r^7 M^4 > 0$, this means that $F_{11}>0$. For completeness, let us mention that as we saw BPS particles satisfy also the repulsive condition (and by the BPS condition the second factor cannot be negative), as well as the ones having $M^2 \leq \frac{2}{\alpha'}\boldsymbol{p}_L^2$, which given the mass formula $\frac{\alpha'}{4}M^2 = \frac{1}{2} \boldsymbol{p}_L^2+N-1$ means that $N=0,1$  \cite{3}.

\end{document}